\documentclass[epj]{svjour}

\usepackage{graphics}

\begin{document}
\title{\boldmath Radiative $\pi^0$ photoproduction on protons in the
       $\Delta^+(\mbox{1232})$ region\unboldmath}

\author{S.~Schumann    \inst{1}\fnmsep\inst{2}\fnmsep
       \thanks{\emph{E-mail address: }\texttt{schumans@kph.uni-mainz.de}},
B.~Boillat             \inst{3},
E.J.~Downie            \inst{1}\fnmsep\inst{4},
P.~Aguar-Bartolom\'{e} \inst{1},
J.~Ahrens              \inst{1},
J.R.M.~Annand          \inst{4},
H.J.~Arends            \inst{1},
R.~Beck                \inst{1}\fnmsep\inst{2},
V.~Bekrenev            \inst{5},
A.~Braghieri           \inst{6},
D.~Branford            \inst{7},
W.J.~Briscoe           \inst{8},
J.W.~Brudvik           \inst{9},
S.~Cherepnya           \inst{10},
R.~Codling             \inst{4},
P.~Drexler             \inst{11},
L.V.~Fil'kov           \inst{10}
D.I.~Glazier           \inst{7},
R.~Gregor              \inst{11},
E.~Heid                \inst{1},
D.~Hornidge            \inst{12},
O.~Jahn                \inst{1},
V.L.~Kashevarov        \inst{10},
R.~Kondratiev          \inst{13},
M.~Korolija            \inst{14},
M.~Kotulla             \inst{11},
D.~Krambrich           \inst{1},
B.~Krusche             \inst{3},
M.~Lang                \inst{1}\fnmsep\inst{2},
V.~Lisin               \inst{13},
K.~Livingston          \inst{4},
S.~Lugert              \inst{11},
I.J.D.~MacGregor       \inst{4},
D.M.~Manley            \inst{15},
M.~Martinez-Fabregate  \inst{1},
J.C.~McGeorge          \inst{4},
D.~Mekterovic          \inst{14},
V.~Metag               \inst{11},
B.M.K.~Nefkens         \inst{9},
A.~Nikolaev            \inst{1}\fnmsep\inst{2},
R.~Novotny             \inst{11},
M.~Ostrick             \inst{1},
R.O.~Owens             \inst{4},
P.~Pedroni             \inst{6},
A.~Polonski            \inst{13},
S.N.~Prakhov           \inst{9},
J.W.~Price             \inst{9},
G.~Rosner              \inst{4},
M.~Rost                \inst{1},
T.~Rostomyan           \inst{6},
D.~Sober               \inst{16},
A.~Starostin           \inst{9},
I.~Supek               \inst{14},
C.M.~Tarbert           \inst{7},
A.~Thomas              \inst{1},
M.~Unverzagt           \inst{1}\fnmsep\inst{2},
Th.~Walcher            \inst{1},
D.P.~Watts             \inst{7}
\and F.~Zehr           \inst{3}
\\\\(Crystal Ball at MAMI, TAPS and A2 Collaborations)
}

\institute{Institut f\"ur Kernphysik,
           Johannes Gutenberg-Universit\"at Mainz, Mainz, Germany
\and Helmholtz-Institut f\"ur Strahlen- und Kernphysik,
     Rheinische Friedrich-Wilhelms-Universit\"at Bonn, Bonn, Germany
\and Institut f\"ur Physik,
     Universit\"at Basel, Basel, Switzerland
\and Department of Physics and Astronomy,
     University of Glasgow, Glasgow, United Kingdom
\and Petersburg Nuclear Physics Institute,
     Gatchina, Russia
\and INFN Sezione di Pavia,
     Pavia, Italy
\and School of Physics,
     University of Edinburgh, Edinburgh, United Kingdom
\and Center for Nuclear Studies,
     The George Washington University, Washington, D.C., USA
\and University of California at Los Angeles,
     Los Angeles, California, USA
\and Lebedev Physical Institute,
     Moscow, Russia
\and II. Physikalisches Institut,
     Justus Liebig-Universit\"at Gie\ss en, Gie\ss en, Germany
\and Mount Allison University,
     Sackville, NB, Canada
\and Institute for Nuclear Research,
     Moscow, Russia
\and Rudjer Boskovic Institute,
     Zagreb, Croatia
\and Kent State University,
     Kent, Ohio, USA
\and The Catholic University of America,
     Washington, D.C., USA
}

\date{Received: date / Revised version: date}

\abstract{
The reaction $\gamma p \rightarrow p \pi^0 \gamma^\prime$ has been measured
with the Crystal Ball / TAPS detectors using the energy-tagged photon beam at
the electron accelerator facility MAMI-B. Energy and angular differential
cross sections for the emitted photon $\gamma^\prime$ and angular differential
cross sections for the $\pi^0$ have been determined with high statistics in
the energy range of the $\Delta^+(1232)$ resonance. Cross sections and the
ratio of the cross section to the non-radiative process 
$\gamma p \rightarrow p \pi^0$ are compared to theoretical reaction models,
having the anomalous magnetic moment $\kappa_{\Delta^+}$  as free parameter.
As the shape of the experimental distributions is not reproduced in detail by
the model calculations, currently no extraction of $\kappa_{\Delta^+}$ is
feasible.
}

\titlerunning{Radiative $\pi^0$ photoproduction on protons in the
              $\Delta^+(1232)$ region}
\authorrunning{S.~Schumann \textit{et al.}}
\maketitle

\section{Introduction}
\label{sec_intro}

The $\Delta(1232)$ as a member of the $J^P = 3/2^+$ baryon decuplet acts as
the first and only well-isolated resonance in elastic pion scattering or pion
photproduction experiments. Its static electromagnetic properties,
particularly the magnetic di\-pole moments $\mu_\Delta$, offer important tests
for baryon structure calculations in the nonpertubative domain of QCD. Several
predictions for the magnetic dipole moments of $\Delta(1232)$ isobars have
been obtained (see table \ref{tab_predict}) using constituent quark based
models \cite{preRQM,preChiQSM,preChiBM} as well as lattice QCD calculations
\cite{Lattice1}. Recently, also the chiral extrapolation of latice results for
$\mu_\Delta$ including the next-to-leading nonanalytic (NLNA) structure of 
$\chi$PT was reported in ref. \cite{Lattice2}. Since there are discrepancies
between the different model calculations, experimental results for 
$\Delta(1232)$ magnetic dipole moments are desirable in order to constrain
this observable at physical quark masses.

\begin{table}
\caption{Predictions of several baryon structure calculations for magnetic
         dipole moments of the $\Delta(1232)$ isobars.}
\label{tab_predict}
\begin{center}
\begin{tabular}{lcccc}
\hline\noalign{\smallskip}
Model & $\mu_{\Delta^{++}}/ \mu_N$ & $\mu_{\Delta^{+}}/ \mu_N$
      & $\mu_{\Delta^{ 0}}/ \mu_N$ & $\mu_{\Delta^{-}}/ \mu_N$  \\
\noalign{\smallskip}\hline\noalign{\smallskip}
$\mathrm{SU(6)}$           & $ 5.58$           & $ 2.79$
                           & $\phantom{-}0.00$ & $-2.79$        \\
RQM       \cite{preRQM}    & $ 4.76$           & $ 2.38$
                           & $\phantom{-}0.00$ & $-2.38$        \\
$\chi$QSM \cite{preChiQSM} & $ 5.40$           & $ 2.65$
                           & $-0.09$           & $-2.83$        \\
$\chi$BM  \cite{preChiBM}  & $ 3.59$           & $ 0.75$
                           & $-2.09$           & $-4.93$        \\
LQCD      \cite{Lattice1}  & $ 4.48\pm0.30$    & $ 2.24\pm0.15$
                           & $\phantom{-}0.00$ & $-2.24\pm0.15$ \\
NLNA      \cite{Lattice2}  & $ 4.99\pm0.56$    & $ 2.49\pm0.27$
                           & $\phantom{-}0.06$ & $-2.45\pm0.27$ \\
\noalign{\smallskip}\hline
\end{tabular}
\end{center}
\end{table}

While experimental studies of the $N \rightarrow \Delta$ transition have
provided access to electromagnetic transition moments (\textit{e.g.}
magnetic dipole $\mu_{N\rightarrow\Delta}$ and electric quadrupole 
$Q_{N\rightarrow\Delta}$ transition moments) \cite{E2M1,TransMoments}, the
static electromagnetic properties of the $\Delta(1232)$ itself, like 
$\mu_\Delta$ and $Q_\Delta$, are difficult to measure due to its short
lifetime of about $\tau_\Delta \simeq 10^{-23}$~s. Thus, the experimental
method of spin precession measurements of $\mu$ that have been performed with
high precision for octet baryons ($N$, $\Lambda$, $\Sigma$, $\Xi$) as well as
for the $\Omega^-$ decuplet baryon is not possible for short-lived states like
the $\Delta(1232)$.

\begin{figure}
\begin{center}
\resizebox{0.46\textwidth}{!}
{
\includegraphics{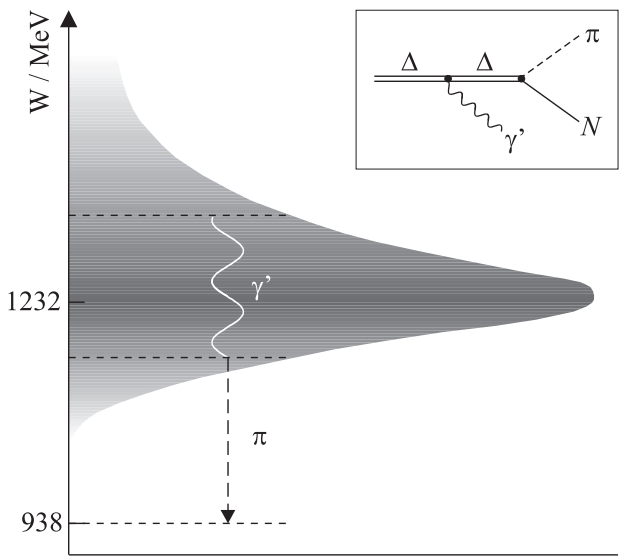}
}
\end{center}
\caption{Method of investigating electromagnetic moments of $\Delta(1232)$
         baryons using a $\Delta\rightarrow \Delta\gamma^\prime$ transition
         within the resonance width and a subsequent decay
         $\Delta \rightarrow N \pi$.}
\label{fig_transition}
\end{figure}

It has been proposed \cite{GammaTrans} to determine the magnetic dipole moment
of the $\Delta(1232)$ from the measurement of electromagnetic transitions
within the resonance width of $\Gamma_\Delta \simeq 120$~MeV. In this process
the nucleon is excited to a $\Delta$ state, which then emits a real photon
$\gamma^\prime$ and subsequently decays into a nucleon and a $\pi$ meson (see
fig. \ref{fig_transition}), hence leading to a $N \pi \gamma^\prime$ final
state. Because of spin and parity conservation, only $M1$, $E2$ and $M3$
multipoles are allowed for this $\Delta \rightarrow \Delta \gamma^\prime$
transition. The amplitude for this process is dominated by magnetic dipole
($M1$) radiation and, therefore, proportional to $\mu_{\Delta}$, as higher
multipole orders give only very small contributions. The electric quadrupole
($E2$) amplitude vanishes in the limit of zero photon energy due to time
reversal symmetry \cite{MDMTheo} and furthermore the $E2/M1$ ratio of about
$-2.5\%$ from the $N\rightarrow \Delta$ transition amplitude indicates that
there is only a very small quadrupole deformation of the $\Delta(1232)$
\cite{E2M1}. Magnetic octupole ($M3$) radiation is suppressed by two
additional powers of photon momentum with respect to the leading $M1$ order.

\begin{figure}
\begin{center}
\resizebox{0.46\textwidth}{!}
{
\includegraphics{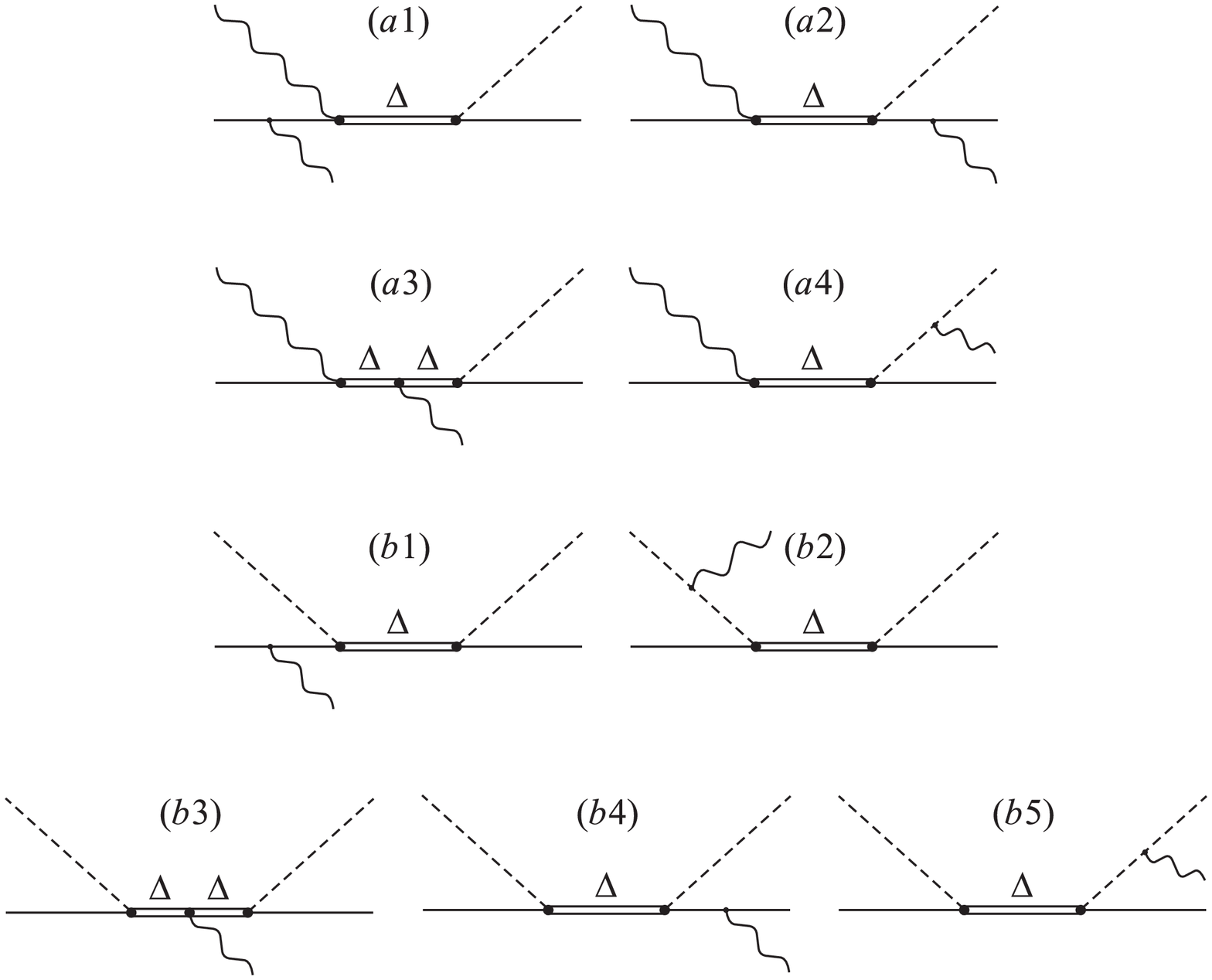}
}
\end{center}
\caption{Bremsstrahlung and $\Delta$-resonant contributions to
         $N \pi \gamma^\prime$ final states for pion photoproduction $(a)$
         and pion scattering $(b)$. Only diagrams $(a3)$ and $(b3)$ are
         sensitive to the magnetic dipole moments $\mu_{\Delta}$.}
\label{fig_diagrams}
\end{figure}

Unfortunately, $N \pi \gamma^\prime$ final states can also result from
bremsstrahlung radiation emitted from incoming and outgoing protons or charged
pions in $\pi N \rightarrow N \pi$ or $\gamma N \rightarrow N \pi$ processes
(see fig. \ref{fig_diagrams}). Such contributions are of the same order as the
$\Delta \rightarrow \Delta \gamma^\prime$ transition and interfere with the
process sensitive to $\mu_{\Delta}$. Thus, any interpretation of experimental
results and determination of the magnetic dipole moment will require and rely
on an accurate theoretical description of all reaction mechanisms contributing
to $ N \pi \gamma^\prime$ final states.

\begin{figure*}
\begin{center}
\resizebox{0.99999\textwidth}{!}
{
\includegraphics{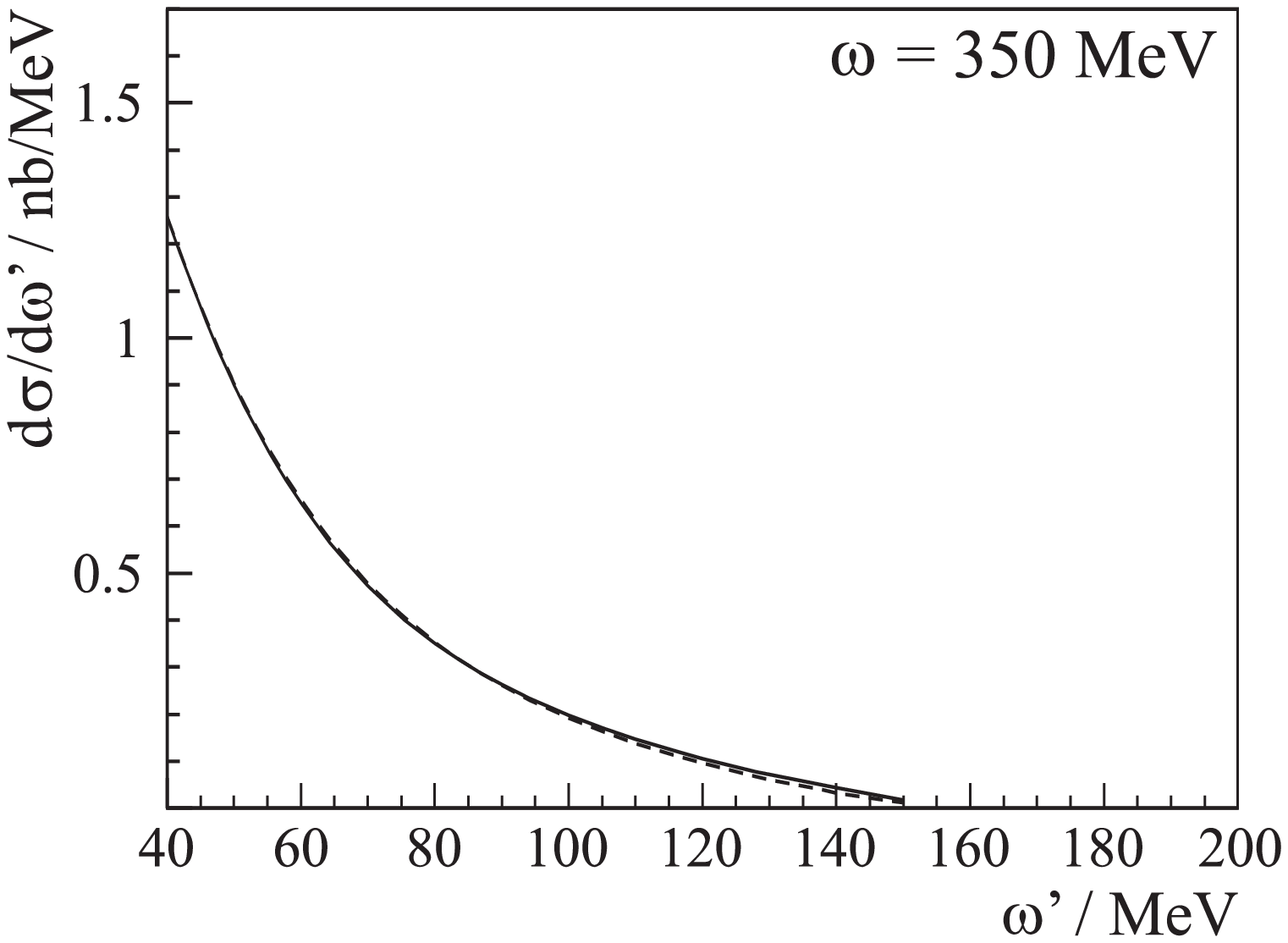}
 \includegraphics{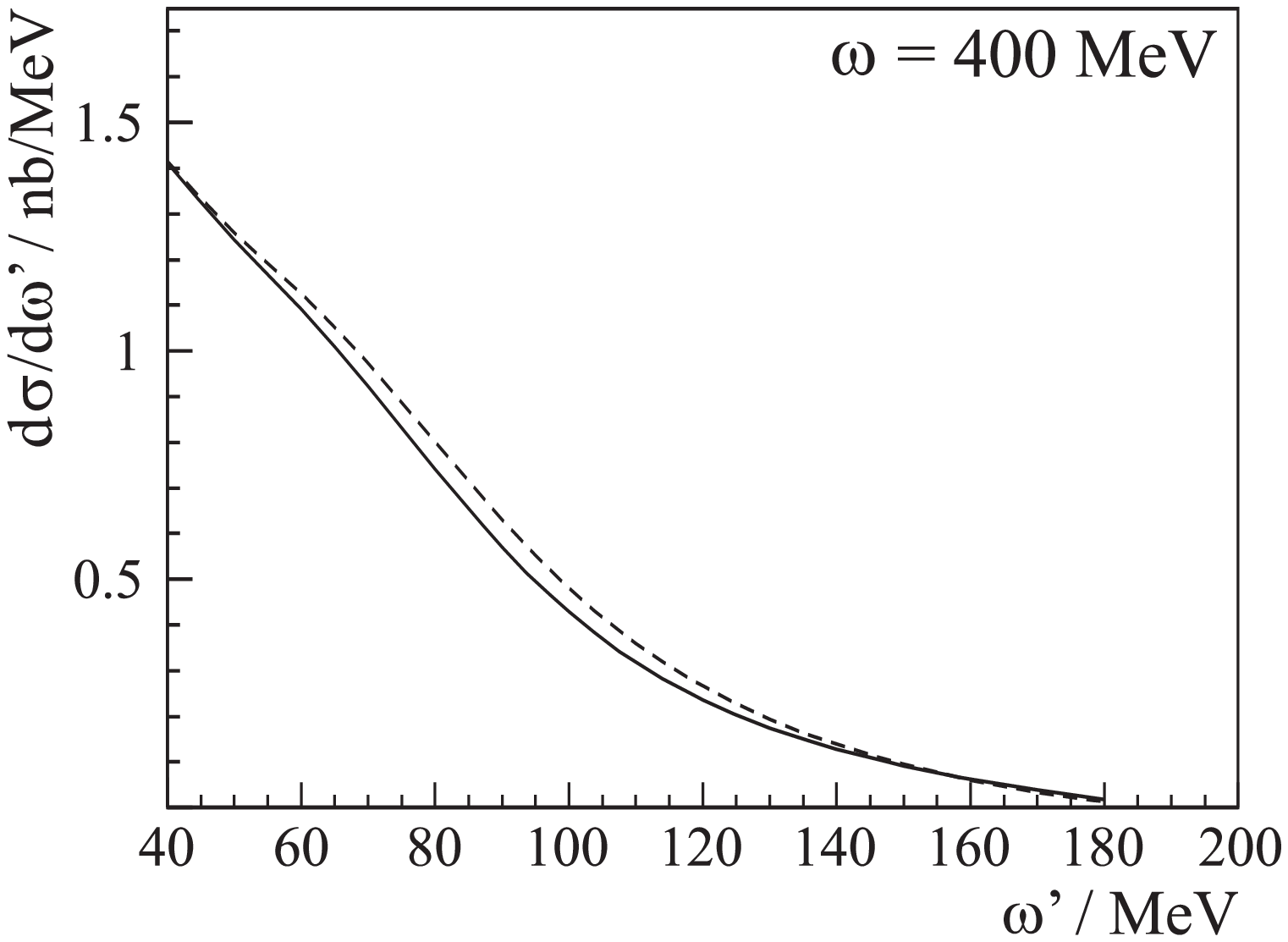}
 \includegraphics{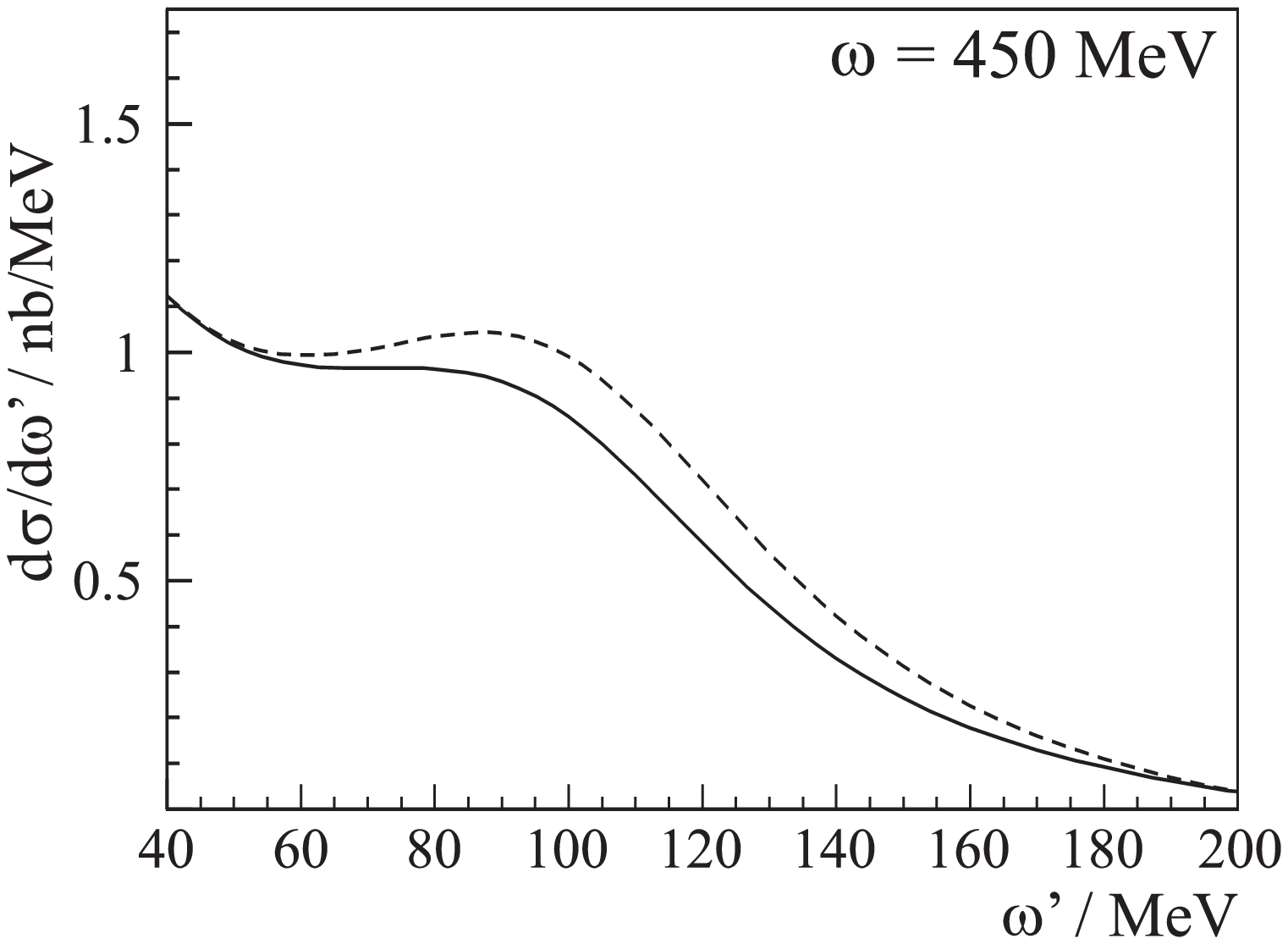}
}
\end{center}
\caption{Model calculations from ref. \cite{MDMTheo} for energy differential
         cross sections $\mathrm d \sigma / \mathrm d \omega^\prime$ of the
         emitted photon $\gamma^\prime$ at different beam energies $\omega$.
         Solid lines are predictions for an anomalous magnetic moment of
         $\kappa_{\Delta^+}=3$, dashed lines are for $\kappa_{\Delta^+}=0$.}
\label{fig_dre01sens}
\end{figure*}

This method for the determination of magnetic dipole moments of unstable
particles was used for the first time in the 1970s at the Lawrence Berkeley
National Laboratory (LBNL) \cite{DeltaPP0,DeltaPP1} and some years later at
the Schweizerisches Institut f\"ur Nuklearforschung / Paul Scherrer Institut
(SIN / PSI) \cite{DeltaPP2,DeltaPP3}. Both experiments used inelastic pion
scattering $\pi^+ p \rightarrow p \pi^+ \gamma^\prime$ in order to extract the
magnetic dipole moment $\mu_{\Delta^{++}}$ of the $\Delta^{++}(1232)$ isobar.
As a result of several theoretical analyses \cite{Theo1,Theo2,Theo3,Theo4} of
both data sets, the Particle Data Group \cite{PDG} quotes a range of
$\mu_{\Delta^{++}} = 3.7$ to $7.5$~$\mu_N$, where $\mu_N = e \hbar/2m_N$ is
the nuclear magneton. These large uncertainties are attributed to strong 
contributions mainly from $\pi^+$, but also from proton brems\-strahlung in
this reaction channel and model dependencies in the theoretical descriptions
of the reaction.

In the case of the $\Delta^+(1232)$ a pioneering measurement of radiative
$\pi^0$ photoproduction $\gamma p \rightarrow p \pi^0 \gamma^\prime$ was
performed by the TAPS / A2 collaborations at MAMI \cite{MDMExp} in 1999. From
this experiment, together with the theoretical description of the reaction
from ref. \cite{MDMTheo}, a value of
\begin{equation}\label{frm_kottu}
\mu_{\Delta^+} = 2.7^{+1.0}_{-1.3}(\mbox{stat})
\pm 1.5(\mbox{syst}) \pm 3.0(\mbox{theo})~\mu_N
\end{equation}
for the magnetic dipole moment of the $\Delta^+(1232)$ was extracted. The
experimental precision of this result is limited by the poor statistics of
around 500 reconstructed $\gamma p \rightarrow p \pi^0 \gamma^\prime$ events
in the previous measurement and its quite large systematic uncertainties,
resulting from an incomplete and inhomogeneous angular acceptance. Also, the
theoretical reaction model from ref. \cite{MDMTheo} that was used to extract
the result (\ref{frm_kottu}) for $\mu_{\Delta^+}$ introduces further
uncertainties due to model dependencies in the description of
$\gamma p \rightarrow p \pi^0 \gamma^\prime$. Thus, a large improvement on the
quality both of experimental data and theoretical descriptions is needed to
get a value for $\mu_{\Delta+}$ that allows discrimination between different
baryon structure calculations. In order to give a quantitative impression of
the experimental and theoretical accuracies needed for the determination of
$\mu_{\Delta^+}$, fig. \ref{fig_dre01sens} shows predictions for energy
differential cross sections $\mathrm d \sigma / \mathrm d \omega^\prime$ for
the emitted photon $\gamma^\prime$ in the reaction 
$\gamma p \rightarrow p \pi^0 \gamma^\prime$. These are evaluated within the
effective Lagrangian framework of ref. \cite{MDMTheo} using different values
for the anomalous magnetic moment $\kappa_{\Delta^+}$. From these curves it
can be estimated that even in the favourable case at $\omega = 450$~MeV and
assuming a precise model description of contributing reaction mechanisms, a
relative experimental precision of about 10\% is needed to get a determination
of the magnetic dipole moment $\mu_{\Delta^+}$ with an accuracy of
approximately $0.5$ to $0.8~\mu_N$, depending on details regarding the
sensitivity to $\mu_{\Delta^+}$ within a particular model framework.

This article presents new data on radiative $\pi^0$ photoproduction in the
$\Delta^+(1232)$ energy region, measured with the Crystal Ball / TAPS detector
set-up and the energy-tagged photon beam at the MAMI-B accelerator facility in
Mainz. This detector system is described in sect. \ref{sec_setup} and offers a
much larger acceptance and efficiency than the TAPS photon spectrometer used
in the previous $\gamma p \rightarrow p \pi^0 \gamma^\prime$ experiment. The
data analysis and identification of
$\gamma p \rightarrow p \pi^0 \gamma^\prime$ reactions are presented in sect.
\ref{sec_analysis} together with a discussion of background contributions from
$\pi^0$ and $\pi^0\pi^0$ production and sect. \ref{sec_models} gives an
overview of different theoretical reaction models for radiative $\pi^0$
photoproduction. In sect. \ref{sec_results} the new experimental results for
various cross sections and other observables are discussed and compared to
predictions of these model calculations. Finally, our conclusions and a
discussion of possible future developments are given in sect.
\ref{sec_outlook}.

\section{Experimental set-up}
\label{sec_setup}

\begin{figure}
\begin{center}
\resizebox{0.46\textwidth}{!}
{
  \includegraphics{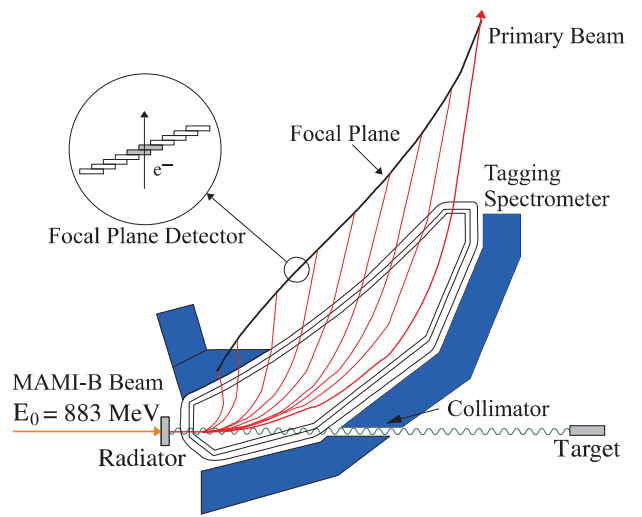}
}
\end{center}
\caption{Glasgow photon tagging system at MAMI-B, producing a
         quasi-monochromatic energy-tagged photon beam.}
\label{fig_tagger}
\end{figure}

The reaction $\gamma p \rightarrow p \pi^0 \gamma^\prime$ has been measured at
the Mainz Microtron (MAMI) electron accelerator facility \cite{MAMI1} using
the Glasgow tagging spectrometer \cite{Tagger1,Tagger2} and the Crystal
Ball / TAPS detector set-up. Bremsstrahlung photons are produced by scattering
the 883 MeV MAMI-B electron beam on a 100~$\mu$m thick diamond radiator, while
scattered electrons are separated from the main beam and momentum-analysed by
a magnetic dipole spectro\-meter and a focal plane detector system made of 353
half-overlapping plastic scintillators (see fig. \ref{fig_tagger}). With the
known beam energy $E_0$ and the energy $E_e$ of scattered electrons the emitted
photon energy $\omega$ is given by
\begin{equation}
\omega = E_0 - E_e
\end{equation}
The resulting energy-tagged quasi-monochromatic photon beam covers an energy
range from 208 to 820~MeV at an average energy resolution of 
$\Delta \omega \simeq 2$~MeV and a tagged photon flux of 
$2.8\cdot 10^7$~s$^{-1}$. The photon flux is determined with an accuracy of
about 5\% by counting the scattered post-bremsstrahlung electrons with the
focal plane detectors of the tagging system and correcting for the loss of
emitted photons due to collimation. The probability for bremsstrahlung photons
 to reach the target (``tagging efficiency'') is periodically measured by a
total-absorption lead glass counter, which is moved into the photon beam line
at reduced beam intensity. At 883~MeV electron beam energy and with a 3~mm
diameter collimator, the tagging efficiency is approximately 34\%.

The photon beam impinges on the target cell (length 4.76~cm) filled with
liquid hydrogen and located in the centre of the Crystal Ball (CB) detector
\cite{CB1,CB2}. The Crystal Ball consists of 672 optically isolated NaI(Tl)
crystals, each read out by an individual photomultiplier tube. Every crystal
has the shape of a truncated triangular pyramid and is about 40.6~cm long,
corresponding to 15.7 radiation lengths. The Crystal Ball covers the full
azimuthal range and a polar angle range from 20$^\circ$ to 160$^\circ$,
resulting in a solid angle coverage of 93\% of $4\pi$. Electromagnetic showers
are reconstructed with an energy resolution of 
$\sigma_E / E = 0.02/(E/\mbox{GeV})^{1/4}$ and angular resolutions of
$\sigma_\theta = 2^\circ\mbox{ to }3^\circ$ and
$\sigma_\phi = \sigma_\theta/\sin\theta$ \cite{CB2}.

\begin{figure}
\begin{center}
\resizebox{0.46\textwidth}{!}
{
\includegraphics{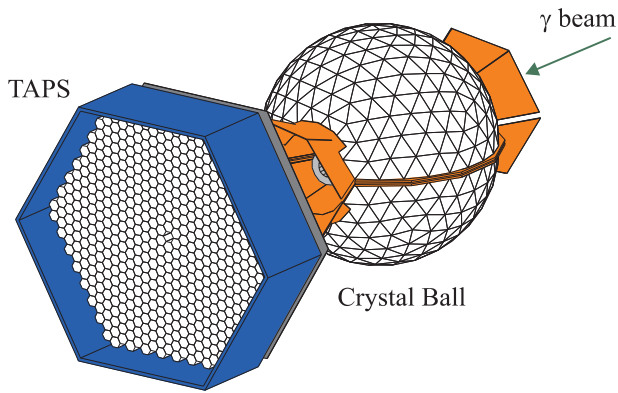}
}
\end{center}
\caption{Crystal Ball / TAPS detector set-up. Additional inner detector
         systems (PID, MWPCs) are installed inside the Crystal Ball beam
         tunnel.}
\label{fig_cbtaps}
\end{figure}

The target is surrounded by the cylindrical Particle Identification Detector
(PID) made of 24 plastic scintillation counters aligned parallel to the beam
axis. Each detector strip is 30~cm long and 2~mm thick. The PID measures the
differential energy loss of charged particles, which  together with the total
energy deposited in the Crystal Ball, is used in a $\Delta E / E$ analysis for
separation of protons and charged pions. To provide precise track
reconstruction for charged particles, two cylindrical wire chambers (MWPCs)
are placed between the PID and the Crystal Ball beam tunnel, covering a polar
angle range from $21^\circ$ to $159^\circ$. Directions of charged particles
are reconstructed from the intersection points of the particle trajectories
with the detection layers in both chambers, where angular resolutions of
$\sigma_\theta=1.3^\circ\mbox{ to }2.3^\circ$ and $\sigma_\phi=1.4^\circ$
and efficiencies of about 95\% for protons and 85\% for $\pi^\pm$ mesons are
achieved. Angular resolutions for reconstructed tracks were determined from
the passage of cosmic radiation through both chambers while efficiencies were
obtained from ana\-lyses of $\gamma p \rightarrow p \pi^0$ and
$\gamma p \rightarrow n \pi^+$ reactions.

Polar angles between 4$^\circ$ and 20$^\circ$ are covered by the TAPS detector
\cite{TAPS1,TAPS2} in a configuration with 510 BaF$_2$ modules arranged as a
hexagonal forward wall at a distance of 1.75~m from the target centre. This
results in a solid angle coverage of the combined Crystal Ball / TAPS detector
set-up (see fig. \ref{fig_cbtaps}) of around 97\% of $4\pi$. Each hexagonally
shaped BaF$_2$ crystal has  an inner diameter of 5.9~cm and a length of 25~cm,
corresponding to 12 radiation lengths. Electromagnetic showers are determined
with an energy resolution of 
$\sigma_E / E = 0.0079/(E/\mbox{GeV})^{1/2} + 0.018$ and angular resolutions
of less than 1$^\circ$ FWHM \cite{TAPS2}. The two components of BaF$_2$
scintillation light (fast component with $\tau_\mathrm{f}= 0.7$~ns, slow
component with $\tau_\mathrm{s}= 620$~ns) are separately digitised with
different ADC gates and can be used in a pulse shape analysis (PSA) for
particle identification. In front of each BaF$_2$ detector crystal a 5~mm
thick plastic scintillator tile acts as a veto detector for charged particles.

With this set-up approximately 800 hours of data were taken in three run
periods during 2004 and 2005. In addition about 120 hours of beam time were
used for background measurements with empty target. Trigger conditions during
all these runs were that the total deposited energy in the Crystal Ball is
more than approximately 60~MeV and a combined sector multiplicity in the
Crystal Ball and TAPS is three or more. For the Crystal Ball such a sector
corresponds to a fixed group of 16 adjacent NaI(Tl) crystals, while TAPS is
divided into four sectors, each consisting of either 127 or 128 BaF$_2$
modu\-les. A sector contributes to the combined multiplicity if the deposited
energy in at least one of its crystals exceeds a threshold of 15~MeV
(Crystal Ball) or 25~MeV (TAPS). Additionally, an independent trigger
condition (downscaled by a factor of 48) requiring a multiplicity of two or
higher was used to record $\gamma p \rightarrow p \pi^0$ events for
consistency checks and calibration purposes.

\section{Data analysis}
\label{sec_analysis}

\subsection{\boldmath Radiative $\pi^0$ photoproduction\unboldmath}

In the beam energy range up to 450~MeV radiative $\pi^0$ production has only
a cross section of approximately $70$~nb which is very small compared to
$\gamma p \rightarrow p \pi^0$ and $\gamma p \rightarrow p \pi^0 \pi^0$
reaction channels with cross sections of between $60$ to $300$~$\mu$b for
single $\pi^0$ and up to $1$~$\mu$b for double $\pi^0$ production. Under some
circumstances these reactions can fake a $p \pi^0 \gamma^\prime$ final state
in the detectors due to either loss of a decay photon in the double $\pi^0$
case or false detector hits in the single $\pi^0$ case. In order to separate
the $\gamma p \rightarrow p \pi^0 \gamma^\prime$ reaction channel from such
background contributions, an exclusive measurement of the
$p \pi^0 \gamma^\prime$ final state has been performed. With measured
4-momenta of all particles, the reaction kinematics are over\-determined,
allowing a variety of kinematic checks.

\begin{figure}
\begin{center}
\resizebox{0.49\textwidth}{!}
{
  \includegraphics{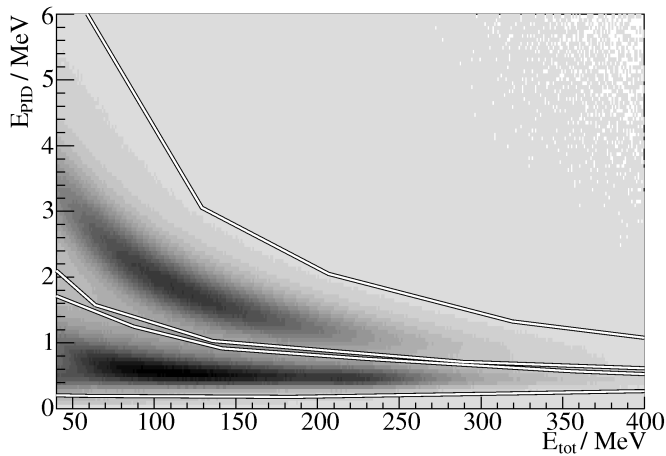}
}
\end{center}
\caption{$\Delta E/E$ analysis for charged particles detected in the Crystal
         Ball, using energy depositions $E_\mathrm{PID}$ and $E_\mathrm{tot}$
         in PID strips and correlated NaI(Tl) clusters. Particles in the upper
         band are accepted as protons, particles in the lower band are
         accepted as $\pi^\pm$ mesons.}
\label{fig_deltaee}
\end{figure}

The first step in the $\gamma p \rightarrow p \pi^0 \gamma^\prime$ analysis
is the selection of events with three neutral clusters and one cluster
identified as a proton. For the Crystal Ball, charged particles are identified
using azimuthal correlations between PID hits and clusters in the NaI(Tl)
array. Further separation of protons and charged pions is done by comparing
the energy loss $E_\mathrm{PID}$ in the PID strip and the total deposited
energy $E_\mathrm{tot}$ from the correlated NaI(Tl) cluster ($\Delta E/E$
technique, see fig. \ref{fig_deltaee}). Precise direction information for
protons detected in the Crystal Ball is obtained from tracks reconstructed
using the MWPC data. Protons detected in the TAPS BaF$_2$ array are identified
using the veto detector hits for separating charged and neutral particles and
the pulse-shape information available from short- and long-gate ADC data.

Measured proton energies are corrected for energy loss in the target and inner
detector systems as well as for different shower propagation compared to
photons in the detector materials. This is done by applying a correction
function obtained from exclusive measurements of
$\gamma p \rightarrow p \pi^0$ reactions, where measured proton energies can
be compared to expected energies calculated from two-body kinematics.

\begin{figure}
\begin{center}
\resizebox{0.49\textwidth}{!}
{
  \includegraphics{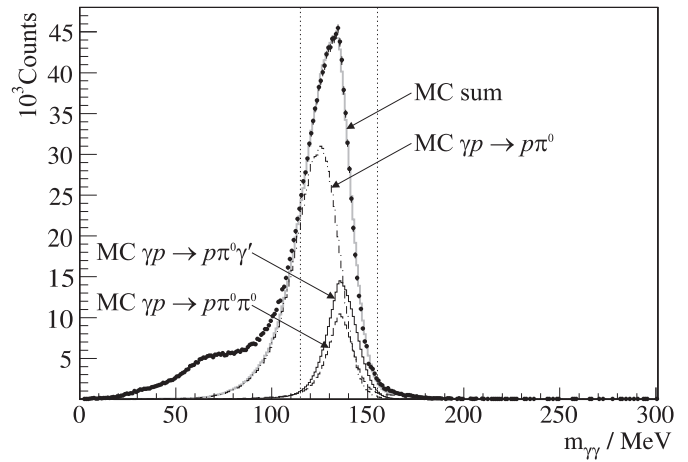}
}
\end{center}
\caption{Invariant $\gamma\gamma$ mass spectrum for the combination giving
         the smallest deviation from the $\pi^0$ mass. Data points represent
         experimental results, while the black lines are from MC simulations
         of $\gamma p \rightarrow p \pi^0$ (dashed-dotted),
         $\gamma p \rightarrow p \pi^0 \gamma^\prime$ (solid) and
         $\gamma p \rightarrow p \pi^0 \pi^0$(dashed), respectively. For
         comparison with experimental data the sum of MC distributions is
         shown (grey solid line). Vertical lines indicate the accepted range
         for reconstructed $\pi^0$ masses.}
\label{fig_invmass}
\end{figure}

The $\pi^0$ meson is identified and reconstructed from its decay into two
photons, using the invariant mass $m_{\gamma \gamma}$ as the selection
criterion. As the reaction $\gamma p \rightarrow p \pi^0 \gamma^\prime$ leads
to a three-photon final state, three different permutations from two out of
these three photons are possible
($\gamma_1 \gamma_2$, $\gamma_1 \gamma_3$, $\gamma_2 \gamma_3$). Therefore,
photons from the combination resulting in an invariant mass with the smallest
deviation from the $\pi^0$ mass are assigned to be $\pi^0$ decay photons, with
the remaining one being interpreted as $\gamma^\prime$. The distribution of
invariant masses for the best photon combination is shown in fig. 
\ref{fig_invmass} together with spectra obtained from MC simulations of the
$\gamma p \rightarrow p \pi^0$, $\gamma p \rightarrow p \pi^0 \gamma^\prime$
and $\gamma p \rightarrow p \pi^0 \pi^0$ reaction channels. The experimental
distribution is dominated by background from non-radiative single $\pi^0$
production, where a secondary particle from an electromagnetic shower
separates and travels some distance in the detector before interacting with
material and creating an additional hit (``split-off'') outside the primary
cluster. In these cases, part of the original photon 4-momentum is not taken
into account, which explains the shift to smaller invariant $\gamma \gamma$
masses in the case of $\gamma p \rightarrow p \pi^0$ background contributions.
The low-energy background contribution at masses 
$m_{\gamma \gamma} < 100$~MeV originates from electromagnetic background
processes from the photon beam and, therefore, is not reproduced by
simulations of $\pi^0$ photoproduction channels. Such contributions are,
however, removed by the condition
$115\mbox{~MeV} < m_{\gamma \gamma} < 155\mbox{~MeV}$ for reconstructed
$\pi^0$ masses.

\begin{figure*}
\begin{center}
\resizebox{0.99999\textwidth}{!}
{
  \includegraphics{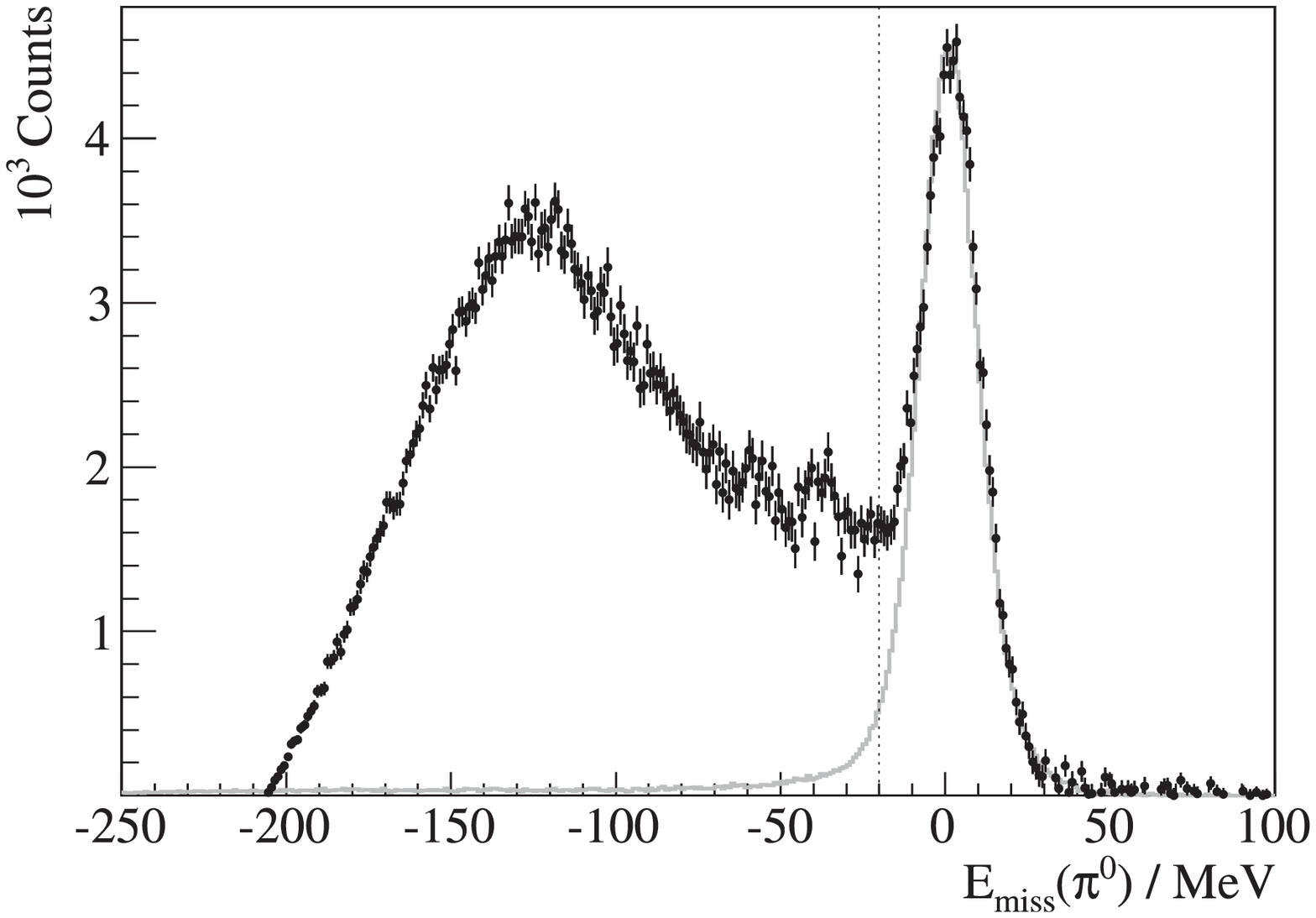}
  \includegraphics{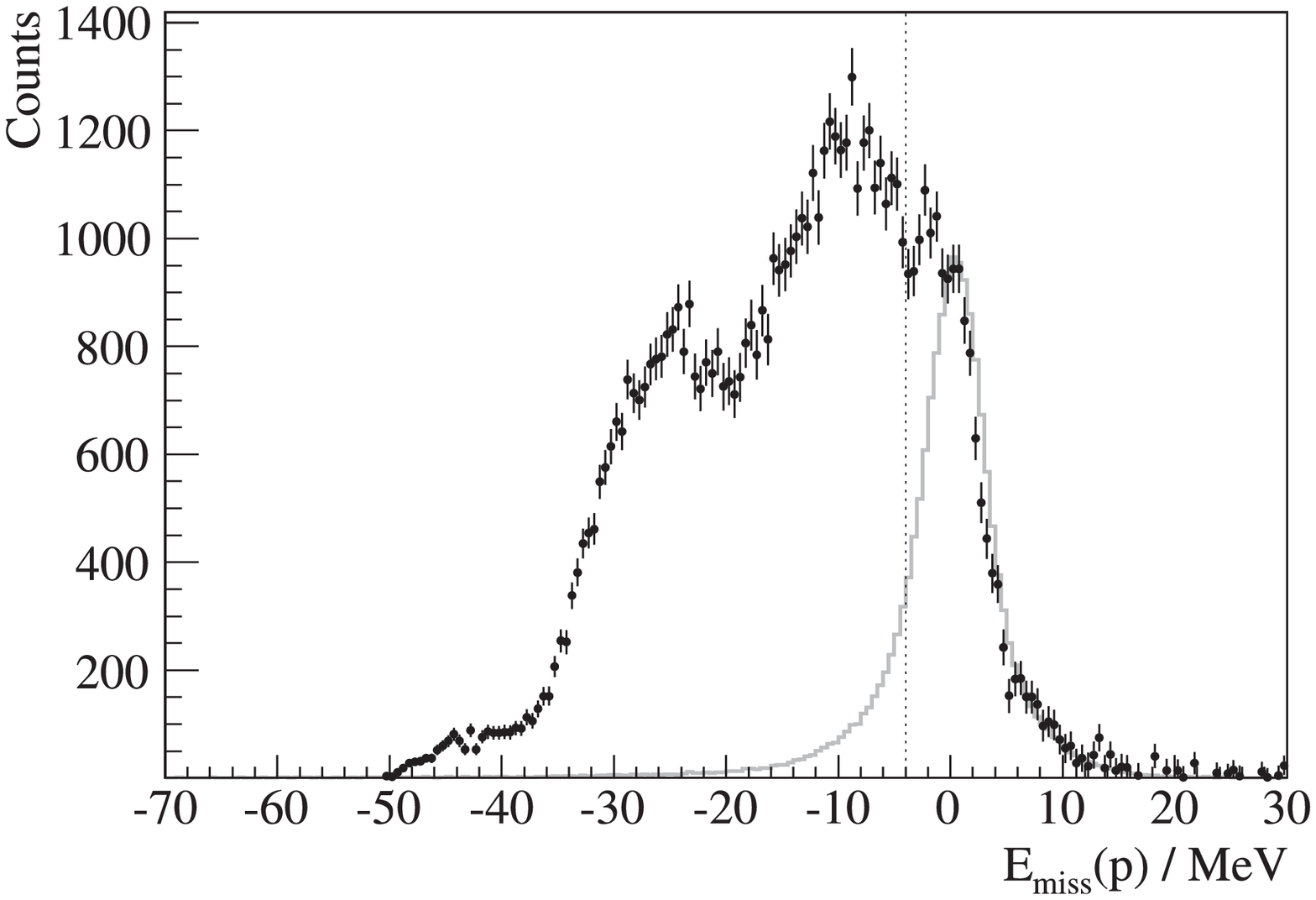}
}
\end{center}
\caption{Missing energy spectra for reconstructed $\pi^0$ mesons (left) and
         protons (right) under the assumption of a
         $\gamma p \rightarrow p \pi^0$ reaction. Data points represent
         results for $\gamma p \rightarrow p \pi^0 \gamma^\prime$ candidates;
         for comparison, the grey solid lines show experimental distributions
         obtained from identified $\gamma p \rightarrow p \pi^0$ events.
         Events resulting in missing energies right of the vertical lines are
         interpreted as $\gamma p \rightarrow p \pi^0$ reactions.}
\label{fig_missenergy}
\end{figure*}

\begin{figure}
\begin{center}
\resizebox{0.49\textwidth}{!}
{
  \includegraphics{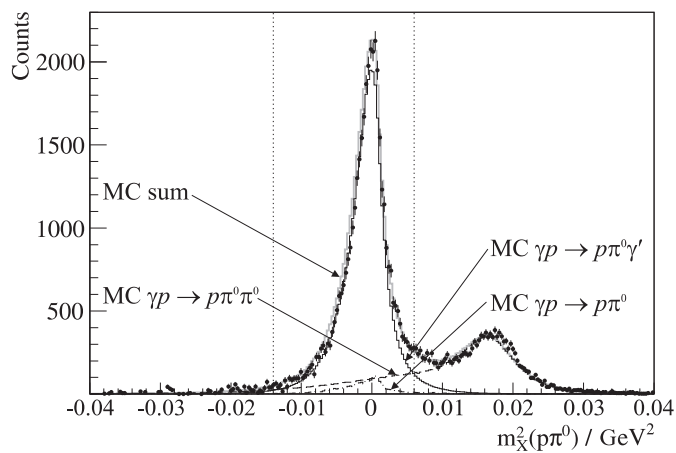}
}
\end{center}
\caption{Missing mass spectrum calculated from the $ p \pi^0$ system. The peak
         around $m_\mathrm{X}^2=0.0182$~GeV$^2$ corresponds to the mass of a
         second $\pi^0$. Data points represent experimental results, while the
         black lines are from MC simulations of $\gamma p \rightarrow p \pi^0$
         (dashed-dotted), $\gamma p \rightarrow p \pi^0 \gamma^\prime$ (solid)
         and $\gamma p \rightarrow p \pi^0 \pi^0$(dashed), respectively. For
         comparison with experimental data the sum of MC distributions is also
         shown (grey solid line). Vertical lines indicate the accepted range
         for missing $\gamma^\prime$ masses.}
\label{fig_missmass2}
\end{figure}

Because of the large background contribution from non-radiative single $\pi^0$
production, candidates for $\gamma p \rightarrow p \pi^0 \gamma^\prime$ events
are explicitly tested for the two-body kinematics of
$\gamma p \rightarrow p \pi^0$ reactions. For the $\pi^0$ meson the missing
energy
\begin{equation}
E_\mathrm{miss}(\pi^0) = E_\pi^* - \frac{s + m_\pi^2 - m_p^2}{2W}
\end{equation}
is calculated, where $E_\pi^*$ denotes the measured $\pi^0$ energy in the
c.m. frame, $m_\pi$ and $m_p$ are the $\pi^0$ and proton masses, respectively,
and
\begin{equation}
W = \sqrt{s} = \sqrt{m_p^2 + 2 \omega m_p}
\end{equation}
is the total c.m. energy at a given photon beam energy $\omega$. Values around
$E_\mathrm{miss}(\pi^0) = 0$~MeV indicate events from
$\gamma p \rightarrow p \pi^0$
reactions, where the measured $\pi^0$ energy corresponds to the expected value
calculated from two-body kinematics (see fig. \ref{fig_missenergy}). An
analogous calculation is done subsequently for the missing energy
$E_\mathrm{miss}(p)$ of the proton, and all events not fulfilling both the
conditions $E_\mathrm{miss}(\pi^0) < -20$~MeV and
$E_\mathrm{miss}(p) < -4$~MeV are interpreted as 
$\gamma p \rightarrow p \pi^0$ background.

\begin{figure}
\begin{center}
\resizebox{0.49\textwidth}{!}
{
  \includegraphics{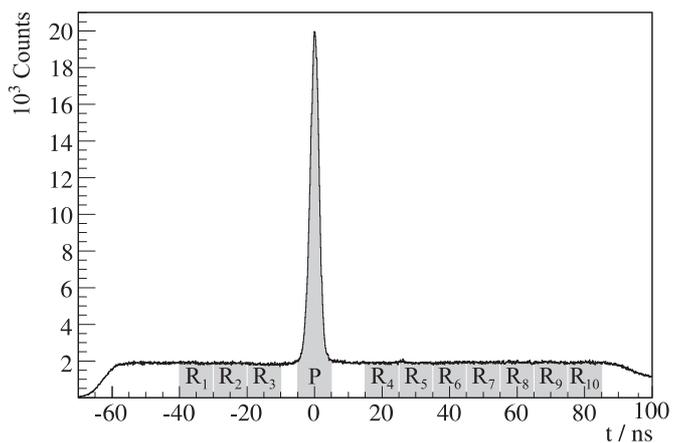}
}
\end{center}
\caption{Time coincidence between Crystal Ball / TAPS and tagging system.
         Random coincidences are evaluated from background windows $R_i$
         and subtracted from events located within the prompt coincidence
         window $P$.}
\label{fig_promptrandom}
\end{figure}

\begin{figure*}
\begin{center}
\resizebox{0.99999\textwidth}{!}
{
  \includegraphics{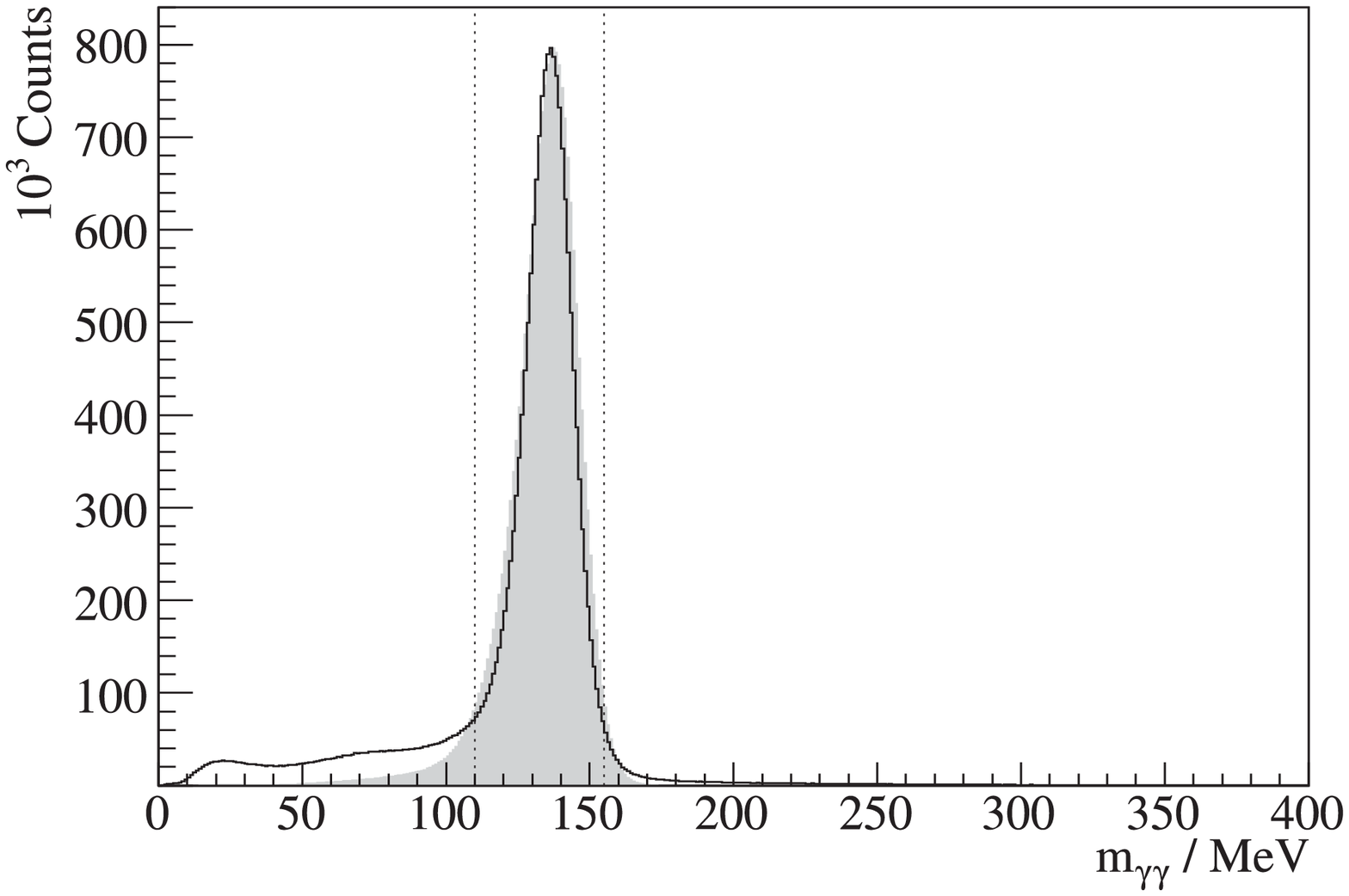}
  \includegraphics{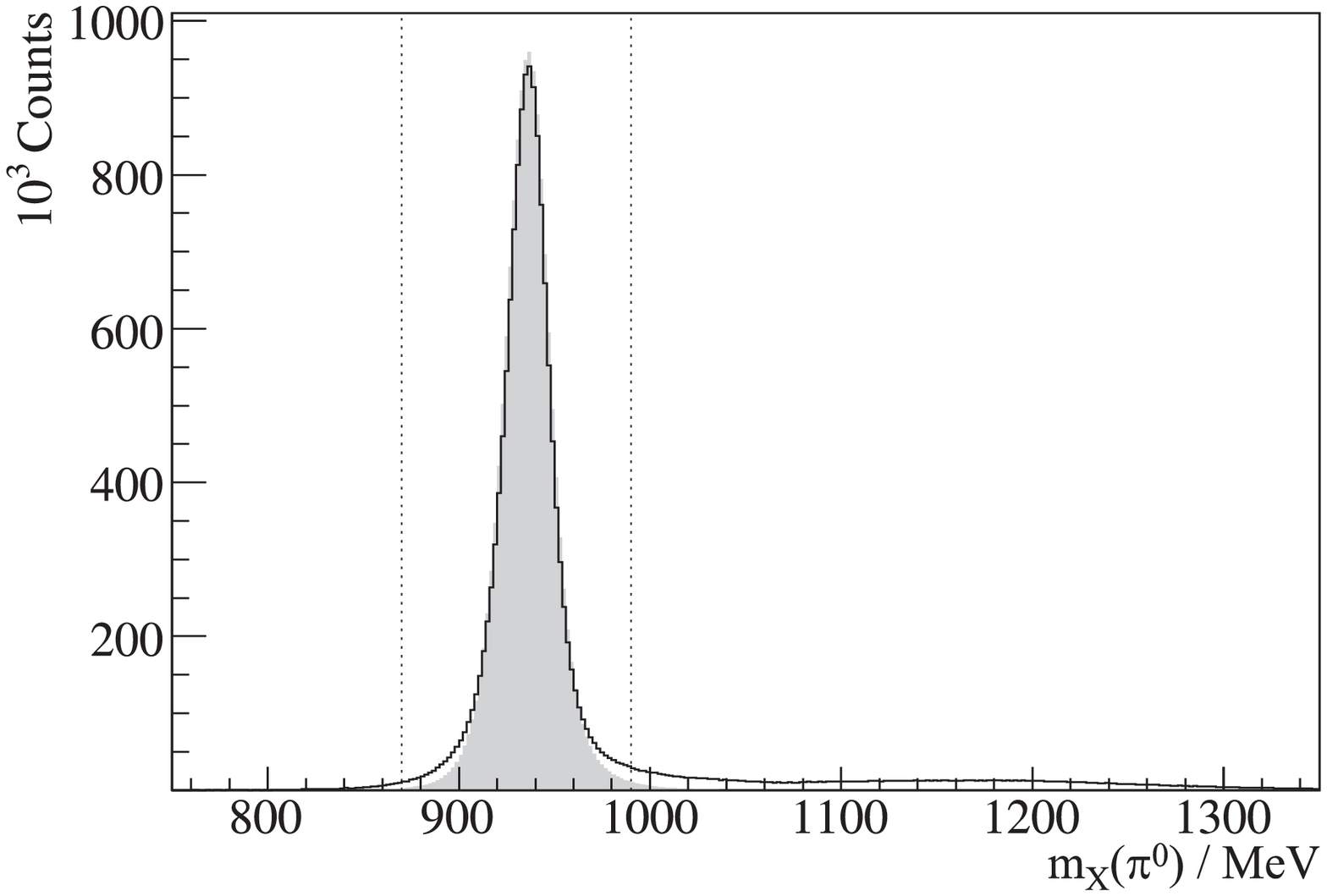}
}
\end{center}
\caption{Analysis of $\gamma p \rightarrow p \pi^0$ reactions. Left: Invariant
         $\gamma\gamma$ mass spectrum for events with two photons. Right:
         Missing mass spectrum calculated from the reconstructed $\pi^0$.
         Solid lines represent experimental results, while the grey shaded
         distributions are from a MC simulation of 
         $\gamma p \rightarrow p \pi^0$ normalised to the experimental peak
         maximum. Vertical lines indicate the ranges of $\pi^0$ and proton
         mass values accepted for further analysis.}
\label{fig_1pi0ana}
\end{figure*}

Double $\pi^0$ production events can lead to three-photon final states if a
decay photon escapes detection either due to the incomplete angular acceptance
or, in case of a highly asymmetric $\pi^0$ decay, the photon energy falls
below the detection threshold of about 25~MeV. Such events are identified
using a missing mass analysis of the detected $p \pi^0$ system without taking
into account the additional photon $\gamma^\prime$. The mass $m_\mathrm{X}$ of
the third final-state particle is calculated from
\begin{equation}
 m_\mathrm{X}^2(p \pi^0)
= \left( \omega + m_p - E_\pi - E_p \right)^2
- \left( \vec k - \vec q_\pi - \vec p_p \right)^2
\end{equation}
where $\omega$, $\vec k$ denote energy and momentum of the beam photon, while
the energies and momenta of proton and $\pi^0$ in the final state are given by
$E_p$, $\vec p_p$ and $E_\pi$, $\vec q_\pi$. For double $\pi^0$ production the
mass of the second $\pi^0$ is reproduced, leading to a peak around
$m_\mathrm{X}^2 = 0.0182$~GeV$^2$ (see fig. \ref{fig_missmass2}), while for
$\gamma p \rightarrow p \pi^0 \gamma^\prime$ the expected value for the
squared missing mass is around $m_\mathrm{X}^2= 0$~GeV$^2$, corresponding to
the mass of the photon $\gamma^\prime$. Thus, events with a squared missing
mass in the range $-0.014 ~\mbox{GeV}^2< m_\mathrm{X}^2 < 0.006~\mbox{GeV}^2$
are accepted as $\gamma p \rightarrow p \pi^0 \gamma^\prime$ reactions.

Finally, remaining random time coincidences bet\-ween the tagging system and
the Crystal Ball / TAPS detector setup are subtracted. These are determined
from background events outside the prompt coincidence peak (see fig.
\ref{fig_promptrandom}). Choosing the same width for prompt and random windows
makes sure that timing conditions are the same for all events. Random
coincidences are evaluated from different time regions $R_{1}$ to $R_{10}$
to get a larger sample of background events and reduce the statistical errors.

\begin{table}
\caption{Background contributions from simulated single and double $\pi^0$
         production reactions to the number of reconstructed events fulfilling
         the analysis conditions for
         $\gamma p \rightarrow p \pi^0 \gamma^\prime$.}
\label{tab_backgnd}
\begin{center}
\begin{tabular}{lccc}
\hline\noalign{\smallskip}
Beam            & Total  & Single $\pi^0$ & Double $\pi^0$ \\
energy $\omega$ & events & background     & background     \\
\noalign{\smallskip}\hline\noalign{\smallskip}
350~MeV & \phantom{1}5880 & 355 & \phantom{111}3 \\
400~MeV &           11996 & 345 & \phantom{11}66 \\
450~MeV &           11839 & 224 &           1139 \\
\noalign{\smallskip}\hline
\end{tabular}
\end{center}
\end{table}

Several \textsc{Geant}~3.21 based MC simulations were performed to determine
detector acceptance and efficiency as well as the remaining background
contributions. Kinematic distributions for 
$\gamma p \rightarrow p \pi^0 \gamma^\prime$ were generated using predictions
from ref. \cite{MDMTheo} for five-fold differential cross sections
$\mathrm{d}^5\sigma / \mathrm{d}\omega^\prime
\mathrm{d}\Omega_{\gamma^\prime} \mathrm{d}\Omega_\pi$
at different beam energies. Simulations using different values of
$\kappa_{\Delta_+}$ were performed to ensure that efficiencies do not depend
on the anomalous magnetic moment used as input for the reaction model. With
these simulations the overall detection and reconstruction efficiency has been
determined to be about 13.5\% for an exclusive measurement of the
$p \pi^0 \gamma^\prime$ final state. The simulation of
$\gamma p \rightarrow p \pi^0$ background reactions is based on MAID 
\cite{MAID1,MAID2} calculations for differential cross sections
$\mathrm{d}\sigma/\mathrm{d}\Omega_\pi$ at different beam energies, while for
$\gamma p \rightarrow p \pi^0 \pi^0$ a phase-space distribution together with
a beam energy dependence according to the total cross section
$\sigma_{\pi\pi}$ from ref. \cite{DblPion1} is used. These simulations show
that the remaining background contributions are approximately 3\% from 
non-radiative single $\pi^0$ and 4\% from double $\pi^0$ production.
Background from $\gamma p \rightarrow p \pi^0 \pi^0$ appears mainly at higher
beam energies around 450~MeV, where it rises up to about 10\%, while at lower
beam energies only a negligible contribution is observed (see table
\ref{tab_backgnd}). The absolute background contribution is derived from
kinematic distributions like fig. \ref{fig_missmass2} and is checked for
consistency with the acceptances for $\gamma p \rightarrow p \pi^0$ and
$\gamma p \rightarrow p \pi^0 \pi^0$ in combination with the particular total
cross sections for both reactions.

For the determination of cross sections the simulated background is subtracted
from the experimental results, leading to a total number of about 27600
reconstructed radiative $\pi^0$ photoproduction events. Cross sections are
calculated from this number of reconstructed 
$\gamma p \rightarrow p \pi^0 \gamma^\prime$ events, corrected for detection
and analysis efficiency and normalised to the number of target protons per
cm$^2$ and the incoming photon beam flux. In order to account for
contributions from the target cell material (125~$\mu$m Kapton) that have been
found to be of the order of 2\%, cross sections are evaluated in the same way
for data taken with an empty target and subsequently subtracted from the full
target results.

\subsection{\boldmath Single and double $\pi^0$ photoproduction\unboldmath}

In addition to the $\gamma p \rightarrow p \pi^0 \gamma^\prime$ analysis cross
sections for non-radiative single and double $\pi^0$ production reactions have
also been extracted from the experimental data. As these processes form the
major background contributions for 
$\gamma p \rightarrow p \pi^0 \gamma^\prime$, a quantitative understanding of
$\gamma p \rightarrow p \pi^0$ and $\gamma p \rightarrow p \pi^0 \pi^0$ is
desirable and can be used to check the consistency of calibrations and data
analysis.

The analysis of single $\pi^0$ production has been limited to events
fulfilling the downscaled trigger condition that the sector multiplicity is
two or more. From this dataset events with two neutral clusters are selected
and the $\pi^0$ is identified and reconstructed using the invariant mass 
$m_{\gamma\gamma}$ of the photon pair (see left panel of fig. 
\ref{fig_1pi0ana}). The identification of the $\gamma p \rightarrow p \pi^0$
reaction channel is then done by evaluating the missing mass
\begin{equation}
m_\mathrm{X}(\pi^0) =
\sqrt{\left(\omega + m_p - E_\pi\right)^2-\left(\vec k - \vec q_\pi\right)^2}
\end{equation}
(see right panel of fig. \ref{fig_1pi0ana}), where $\omega$, $\vec k$ and
$E_\pi$, $\vec q_\pi$ denote energy and momentum of the beam photon and the
reconstructed $\pi^0$, respectively. To determine the total cross section,
random and empty target background is subtracted and the MAID-based MC
simulation of $\gamma p \rightarrow p \pi^0$, as described above, is used to
correct the number of reconstructed events for detection and analysis
efficiency, which is about 67\% for an inclusive measurement of single $\pi^0$
production. The total systematic uncertainty for this reaction is estimated to
be about 6\%, which covers uncertainties of 5\% in photon flux determination
and 1.8\% for target length and density as well as 3\% for acceptance and 
efficiency corrections.

\begin{figure}
\begin{center}
\resizebox{0.49\textwidth}{!}
{
  \includegraphics{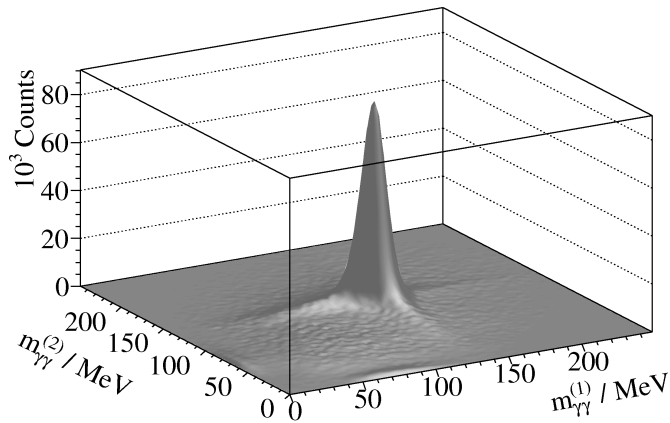}
}
\end{center}
\caption{Invariant masses of $\gamma\gamma$ pairs for double $\pi^0$
         production. Only the best combination with minimal $\chi^2$ according
         to eq. (\ref{frm_ChiSqDbl}) is shown.}
\label{fig_2pi0ana1}
\end{figure}

\begin{figure}
\begin{center}
\resizebox{0.49\textwidth}{!}
{
  \includegraphics{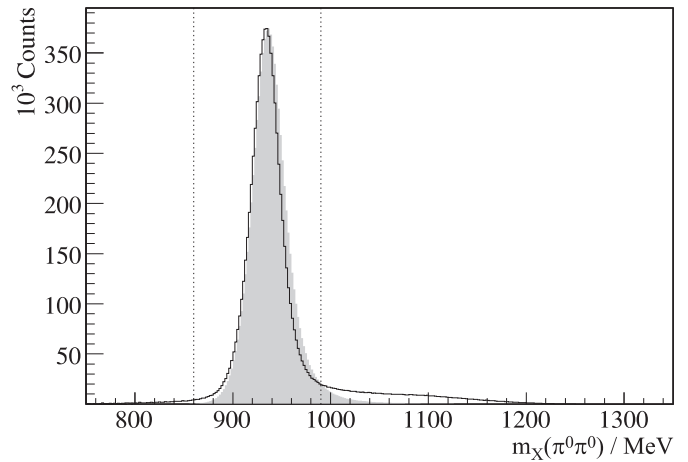}
}
\end{center}
\caption{Missing mass spectrum for double $\pi^0$ production. The solid line
         represents experimental results, while the grey shaded distribution
         is from a MC simulation of $\gamma p \rightarrow p \pi^0 \pi^0$
         normalised to the experimental peak maximum. Vertical lines indicate
         the accepted ranges for missing proton masses.}
\label{fig_2pi0ana2}
\end{figure}

In the case of double $\pi^0$ production events with four detected photons are
selected and the $\pi^0$ mesons are reconstructed from the best combination of
$\gamma\gamma$ pairs. As there are 3 different possible permutations, the
combination with the best simultaneous reconstruction of both pions,
\textit{i.e.} with minimal
\begin{equation}
\label{frm_ChiSqDbl}
\chi^2 = \left(m_{\gamma\gamma}^{(1)} - m_\pi\right)^2 +
         \left(m_{\gamma\gamma}^{(2)} - m_\pi\right)^2
\end{equation}
is selected (see fig. \ref{fig_2pi0ana1}). Similar to the single $\pi^0$ case,
the final identification of the reaction channel
$\gamma p \rightarrow p \pi^0\pi^0$ is done using the missing mass (see fig.
\ref{fig_2pi0ana2}) calculated from the reconstructed $\pi^0$ mesons according
to
\begin{equation}
m_\mathrm{X}(\pi^0\pi^0)
=\sqrt{\left(k + p_\mathrm{T} - q_\pi^{(1)}- q_\pi^{(2)}\right)^2}
\end{equation}
where $k = (\omega, \vec k)$ and $p_\mathrm{T} = (m_p, \vec 0)$ denote the
4-momenta of the initial beam photon and target proton, while the 4-momenta of
the final state pions are given by $q_\pi = (E_\pi, \vec q_\pi)$. After
subtraction of empty target and random background contributions, the total
cross section is calculated using the detection and reconstruction efficiency
of about 42\% for the inclusive case, determined with the phase-space MC
simulation described above. Kinematic discrepancies between simulation and
experimental data are taken into account in the systematic uncertainty of the
efficiency correction, which is estimated to be 5\%. Together with the
absolute normalisation uncertainties from photon flux and target density,
which are the same as for the single $\pi^0$ case, this gives a total
systematic uncertainty of about 7\% for the double $\pi^0$ production cross
sections.

\begin{figure*}
\begin{center}
\resizebox{0.99999\textwidth}{!}
{
  \includegraphics{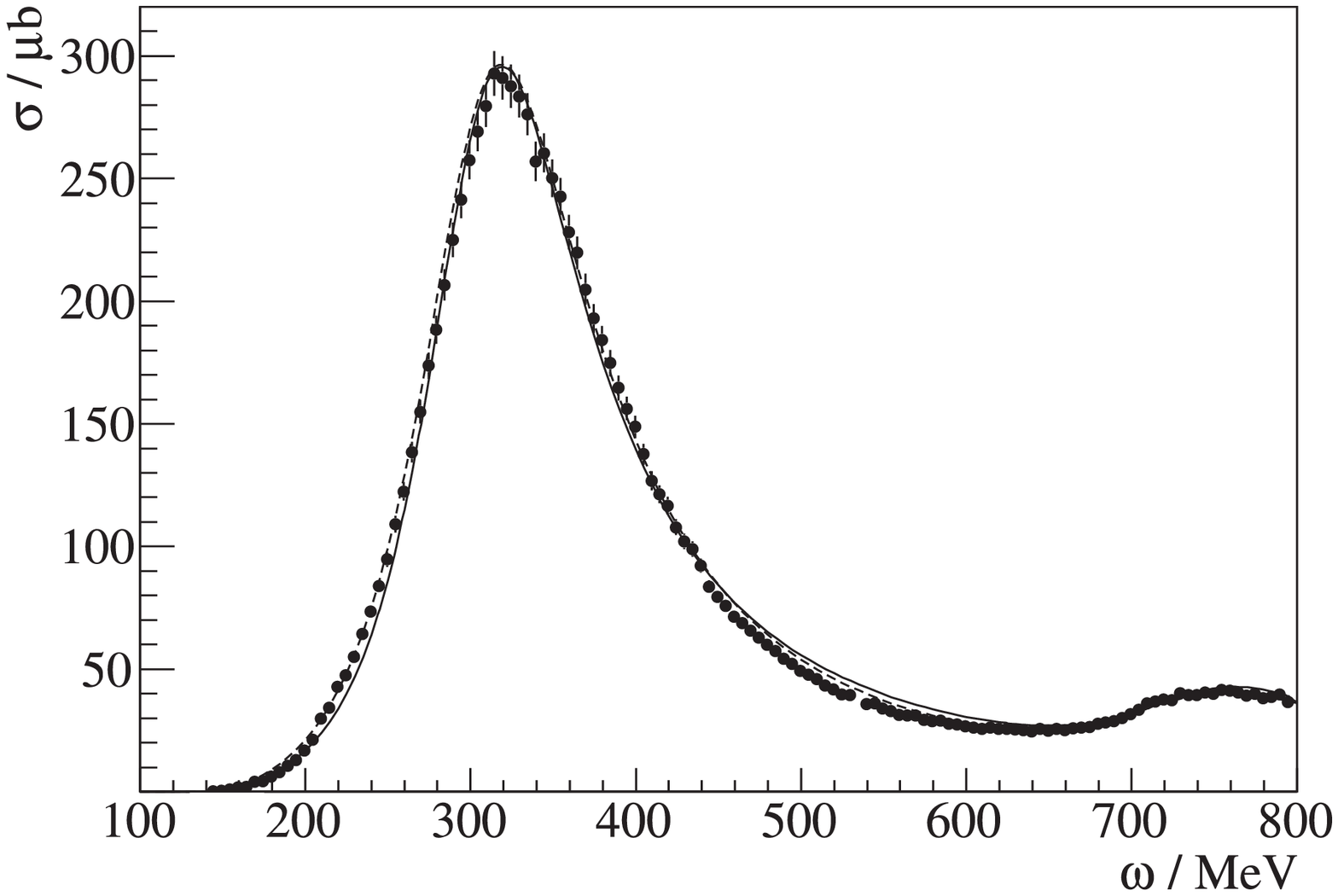}
  \includegraphics{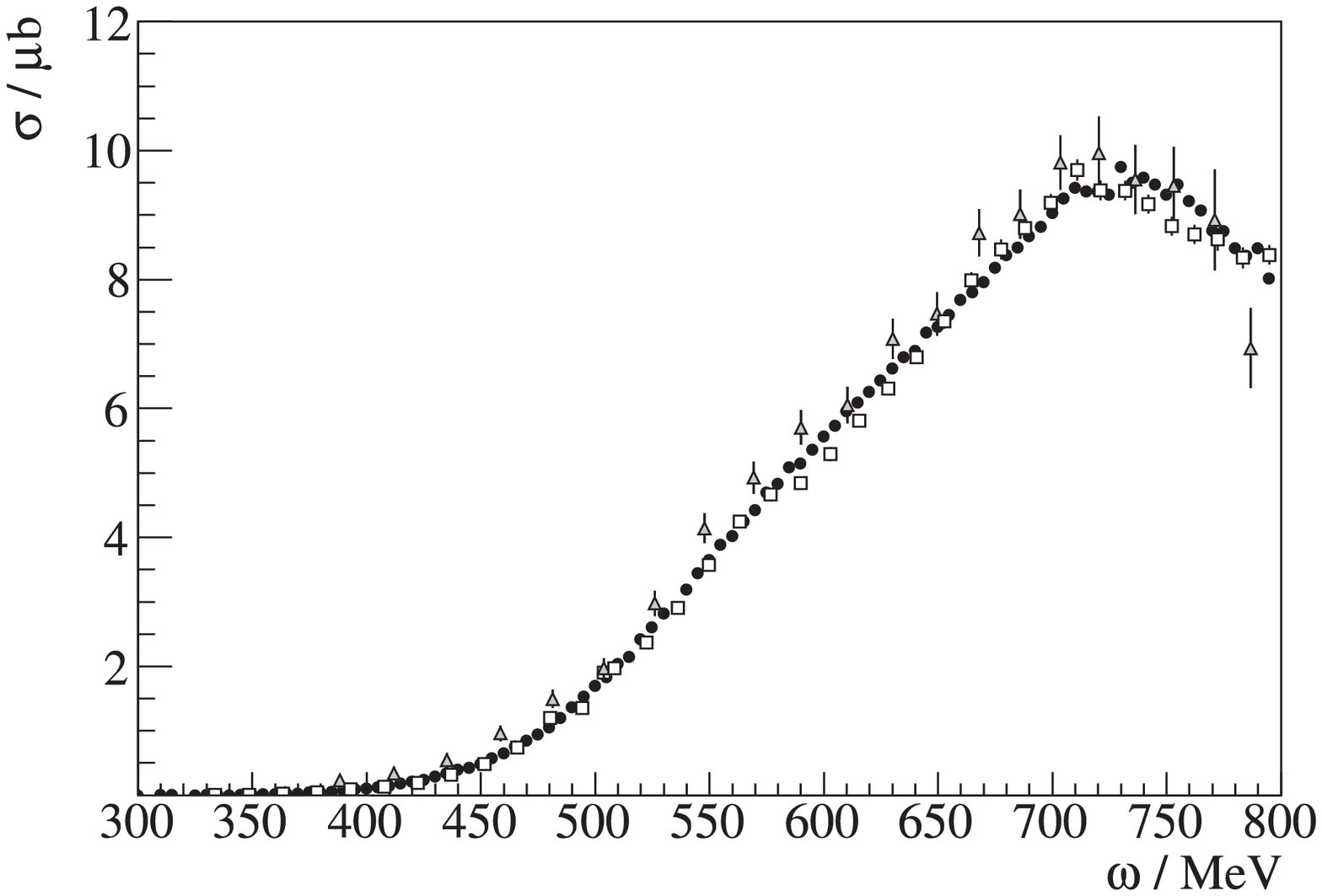}
}
\end{center}
\caption{Total cross sections for background reactions
         $\gamma p \rightarrow p \pi^0$ (left) and 
         $\gamma p \rightarrow p \pi^0 \pi^0$ (right). Black points represent
         Crystal Ball / TAPS results, black lines show MAID \cite{MAID1,MAID2}
         (solid) and SAID \cite{SAID} (dashed) predictions for single $\pi^0$
         production. White squares and grey triangles are double $\pi^0$
         results from refs. \cite{DblPion1a} and \cite{DblPion2},
         respectively. Error bars include both statistical and systematic
         uncertainties for $\gamma p \rightarrow p \pi^0$ and statistical
         uncertainties only for $\gamma p \rightarrow p \pi^0 \pi^0$.}
\label{fig_totalppi0}
\end{figure*}

\section{Reaction models}
\label{sec_models}

In order to determine the $\Delta^+(1232)$ magnetic dipole moment
$\mu_{\Delta^+}$ from experimental results for
$\gamma p \rightarrow p \pi^0 \gamma^\prime$ an accurate theoretical
description of all contributing processes is required. While first theoretical
calculations \cite{Dresonant1,Dresonant2,Dresonant3} considered only the
$\Delta$-resonant mechanism and, therefore, could not reproduce experimental
results, an extended description including several other contributions has
been presented in ref. \cite{MDMTheo}. This model was used for the extraction
of $\mu_{\Delta^+}$ from the previous TAPS / A2 experiment in ref.
\cite{MDMExp}. Its calculations are done within the context of an effective
Lagrangian formalism, where in the first step the $\Delta$-resonant mechanism
as well as a background of non-resonant contributions (Born terms, vector
meson exchange) are taken into account for a tree-level description of the
$\gamma p \rightarrow p \pi^0$ process. Then, the additional photon
$\gamma^\prime$ is coupled in a gauge-invariant way to charged particles,
leading to the $\gamma p \rightarrow p \pi^0 \gamma^\prime$ reaction and
introducing the anomalous magnetic moment $\kappa_{\Delta^+}$ as a new
parameter to the $\gamma\Delta\Delta$ vertex, where $\mu_{\Delta^+}$ and
$\kappa_{\Delta^+}$ are related according to
\begin{equation}\label{frm_kappamu}
 \mu_{\Delta^+}
 = \left(1 + \kappa_{\Delta^+}\right)\frac{e\hbar}{2 m_{\Delta}}
 = \left(1 + \kappa_{\Delta^+}\right)\frac{m_N}{m_{\Delta}}~\mu_N
\end{equation}
with $m_N$ and $m_{\Delta}$ the nucleon and $\Delta^+(1232)$ masses,
respectively.

Since the earlier determination of $\mu_{\Delta^+}$ several improvements have
been made to the theoretical calculations. The effective Lagrangian approach
from ref. \cite{MDMTheo} has been extended in ref. \cite{Unitary} to a
unitarised dynamical model of $\gamma p \rightarrow N \pi \gamma^\prime$
reactions (in the following referred to as ``unitary model''), including
$\pi N$ rescattering effects in an on-shell approximation for intermediate
particles ($K$ ma\-trix approximation). Furthermore these rescattering
contributions are treated in the soft-photon limit
$\omega^\prime \rightarrow 0$ for the final $\gamma^\prime$, where the $T$
matrix for $\gamma p \rightarrow N \pi \gamma^\prime$ reactions is directly
proportional to the full $T$ matrix for $\gamma p \rightarrow N \pi$
processes.

In another approach, descriptions of radiative pion photoproduction
$\gamma p \rightarrow N \pi \gamma^\prime$ using chiral effective field theory
($\chi$EFT) have been developed \cite{ChEFT1,ChEFT2} in order to use a more
consistent and systematic framework compared to the ``phenomenological''
effective Lagrangian models from refs. \cite{MDMTheo} and \cite{Unitary}. The
framework of $\chi$EFT allows the calculation of chiral loop corrections in a
consistent way following quantum field theory, where the chiral symmetry of
the low-energy strong interaction as well as other general principles like
unitarity and analyticity are included to any given order in a systematic
expansion over the energy scales and hadronic degrees of freedom. The latest
calculations in ref. \cite{ChEFT2} are based on a next-to-leading order (NLO)
chiral expansion with $\Delta$-isobar degrees of freedom (using the 
$\delta$-expansion power counting scheme described in refs. 
\cite{deltaExp1,deltaExp2}) and a next-next-to-leading order (NNLO)
soft-photon expansion with respect to the emitted $\gamma^\prime$ energy,
since this is the order at which $\mu_{\Delta^+}$ first appears.

\begin{figure}
\begin{center}
\resizebox{0.49\textwidth}{!}
{
  \includegraphics{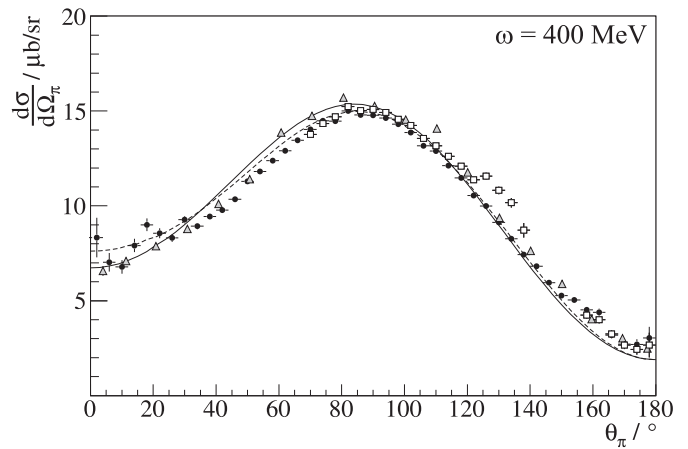}
}
\end{center}
\caption{Differential cross section $\mathrm d\sigma/\mathrm d \Omega_\pi$ for
         $\gamma p \rightarrow p \pi^0$. Black points and white squares
         represent Crystal Ball / TAPS results for inclusive ($\pi^0$ only)
         and exclusive ($p$ and $\pi^0$) measurements, respectively, while
         grey triangles are results from ref. \cite{SglPion1}. Black lines
         show MAID \cite{MAID1,MAID2} (solid) and SAID \cite{SAID} (dashed)
         predictions. Error bars include statistical uncertainties only.}
\label{fig_sgldsdO}
\end{figure}

\begin{figure*}
\begin{center}
\resizebox{0.99999\textwidth}{!}
{
  \includegraphics{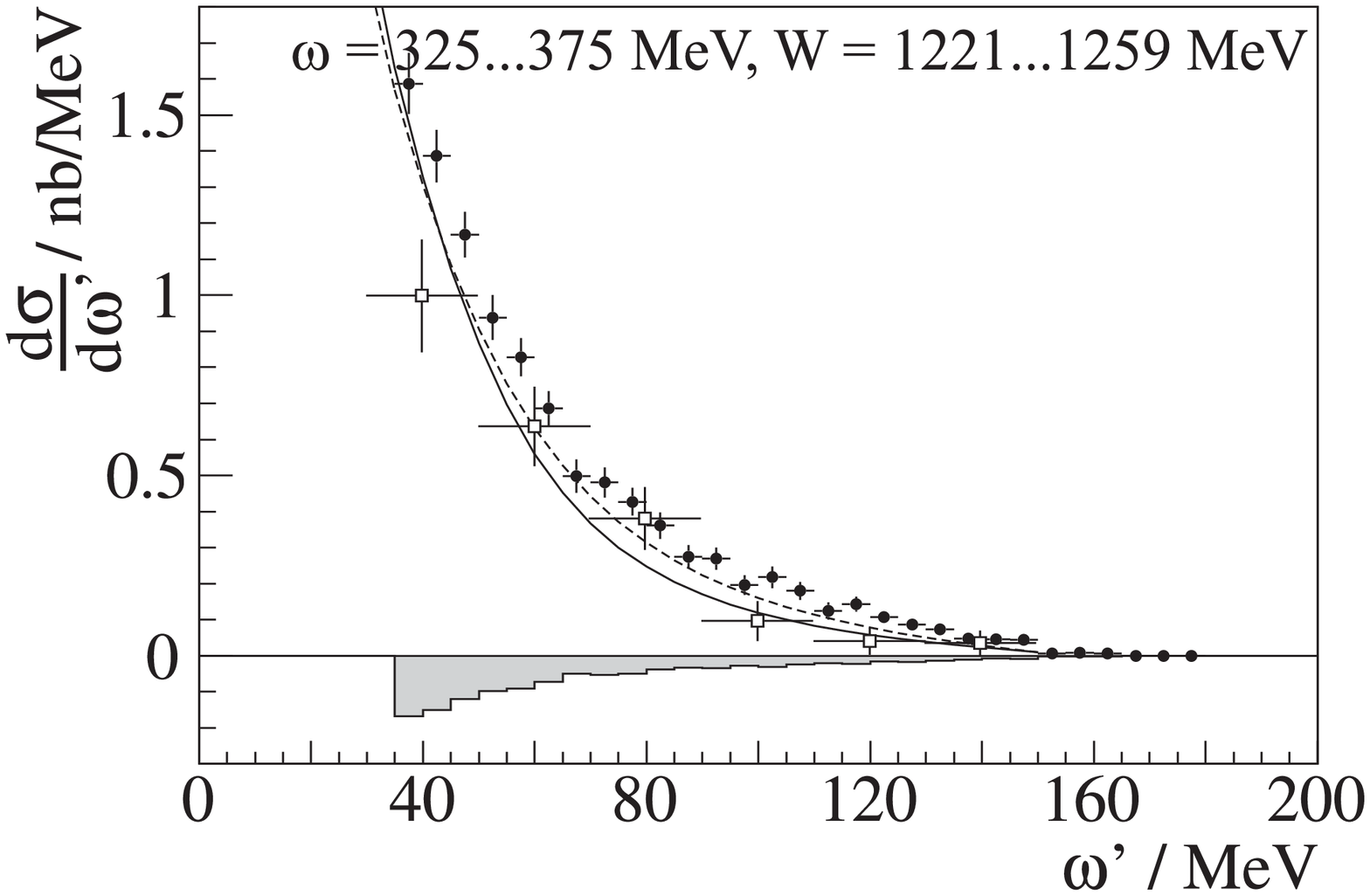}
  \includegraphics{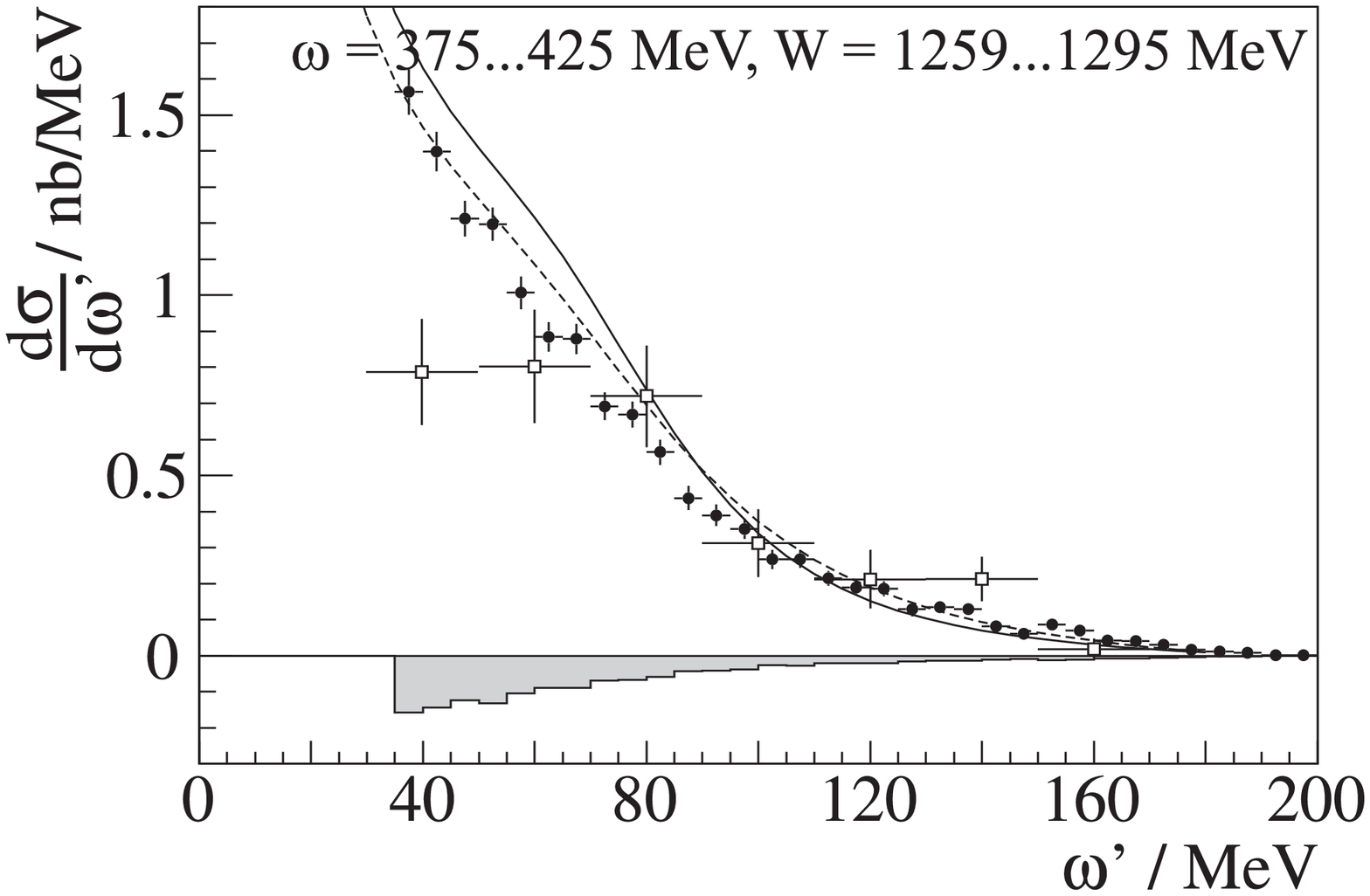}
  \includegraphics{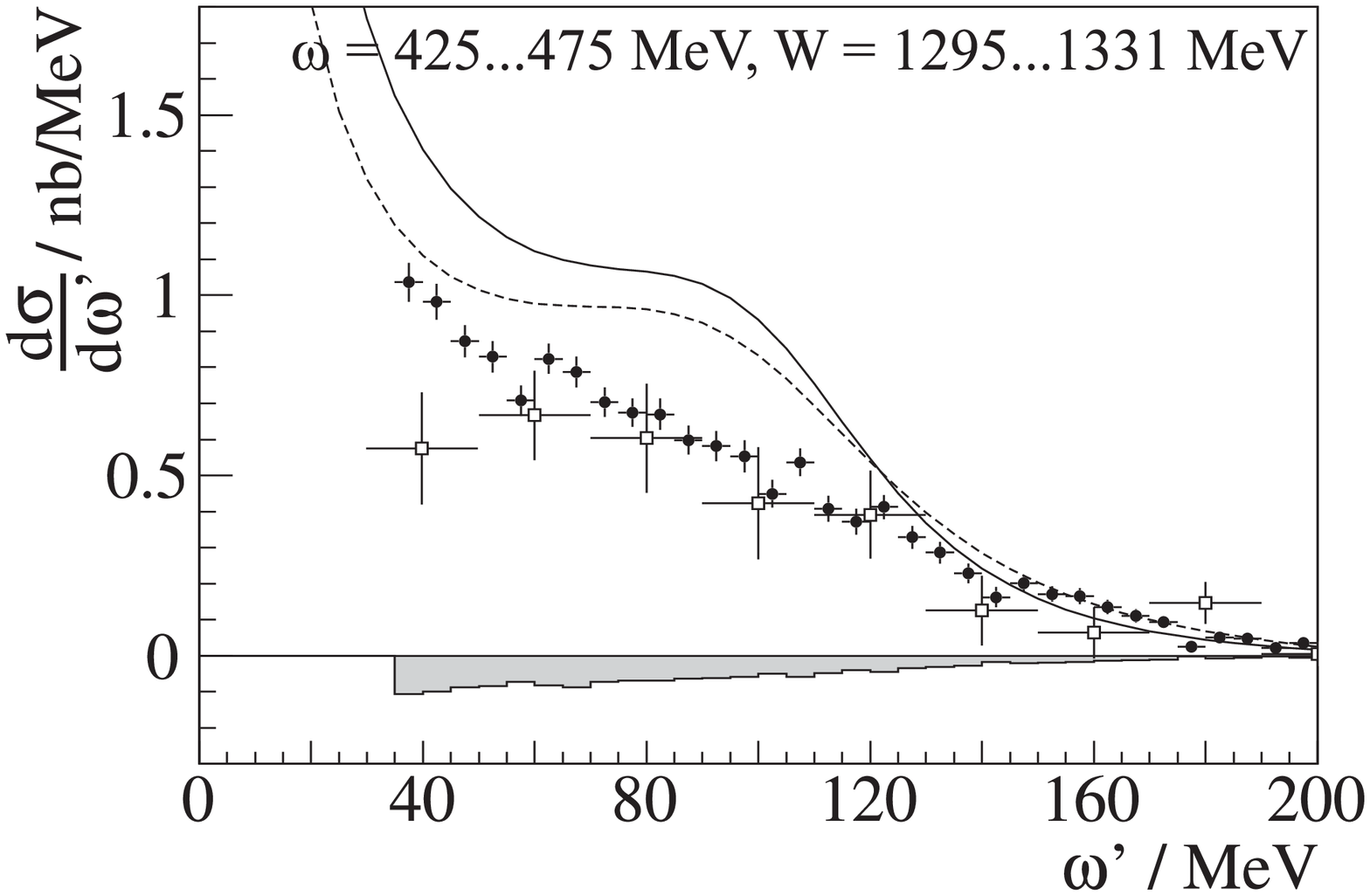}
}

\resizebox{0.99999\textwidth}{!}
{
  \includegraphics{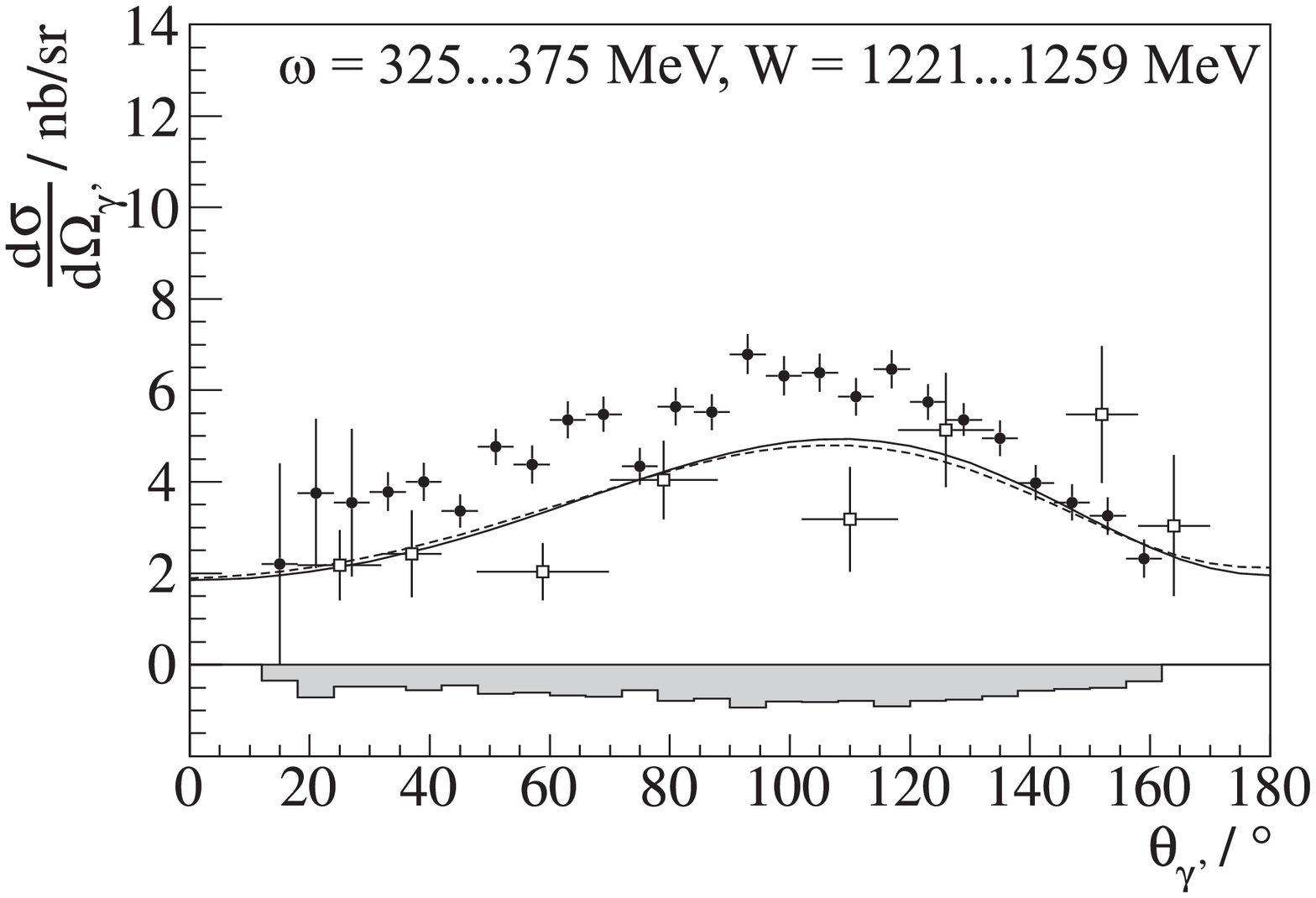}
  \includegraphics{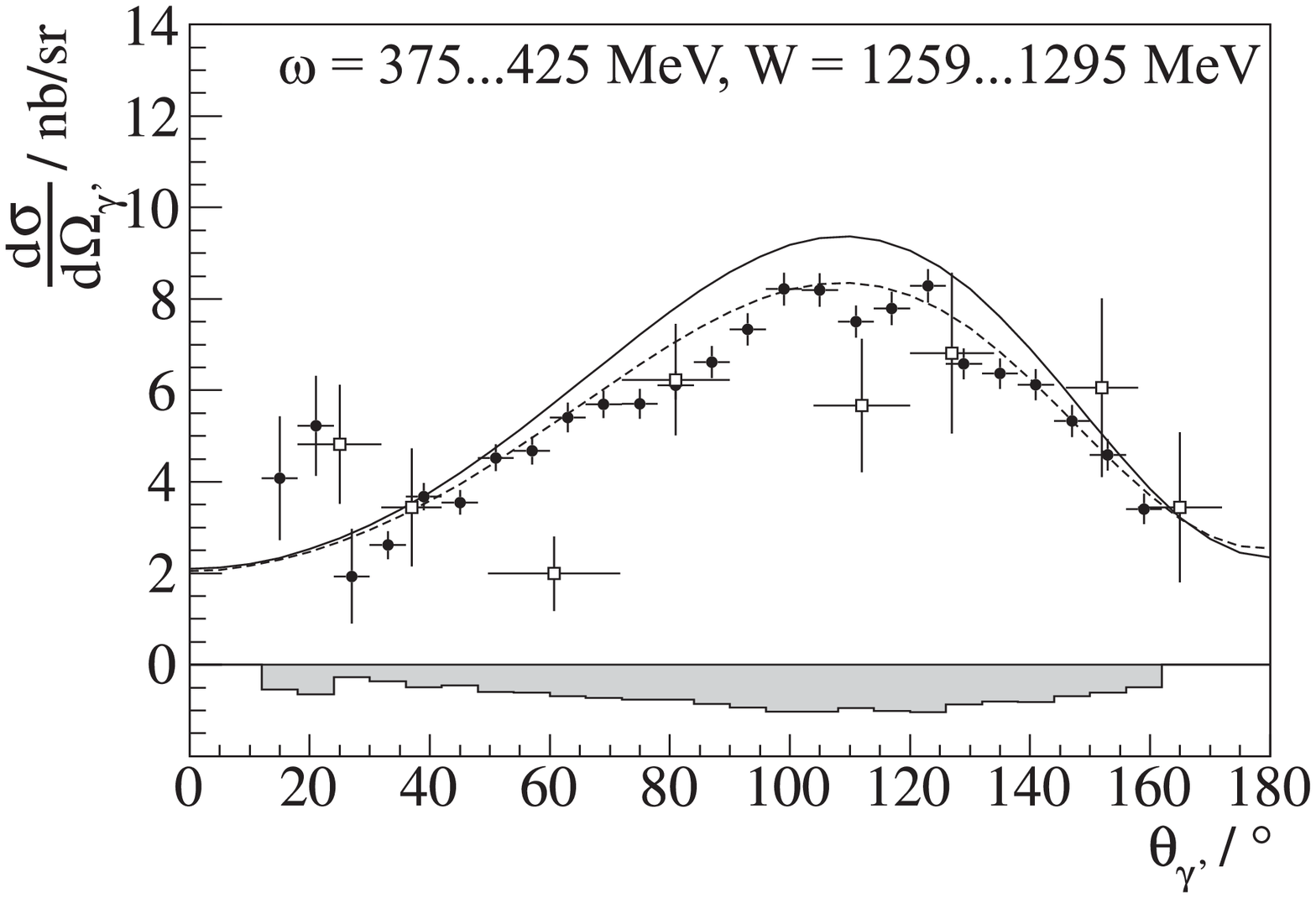}
  \includegraphics{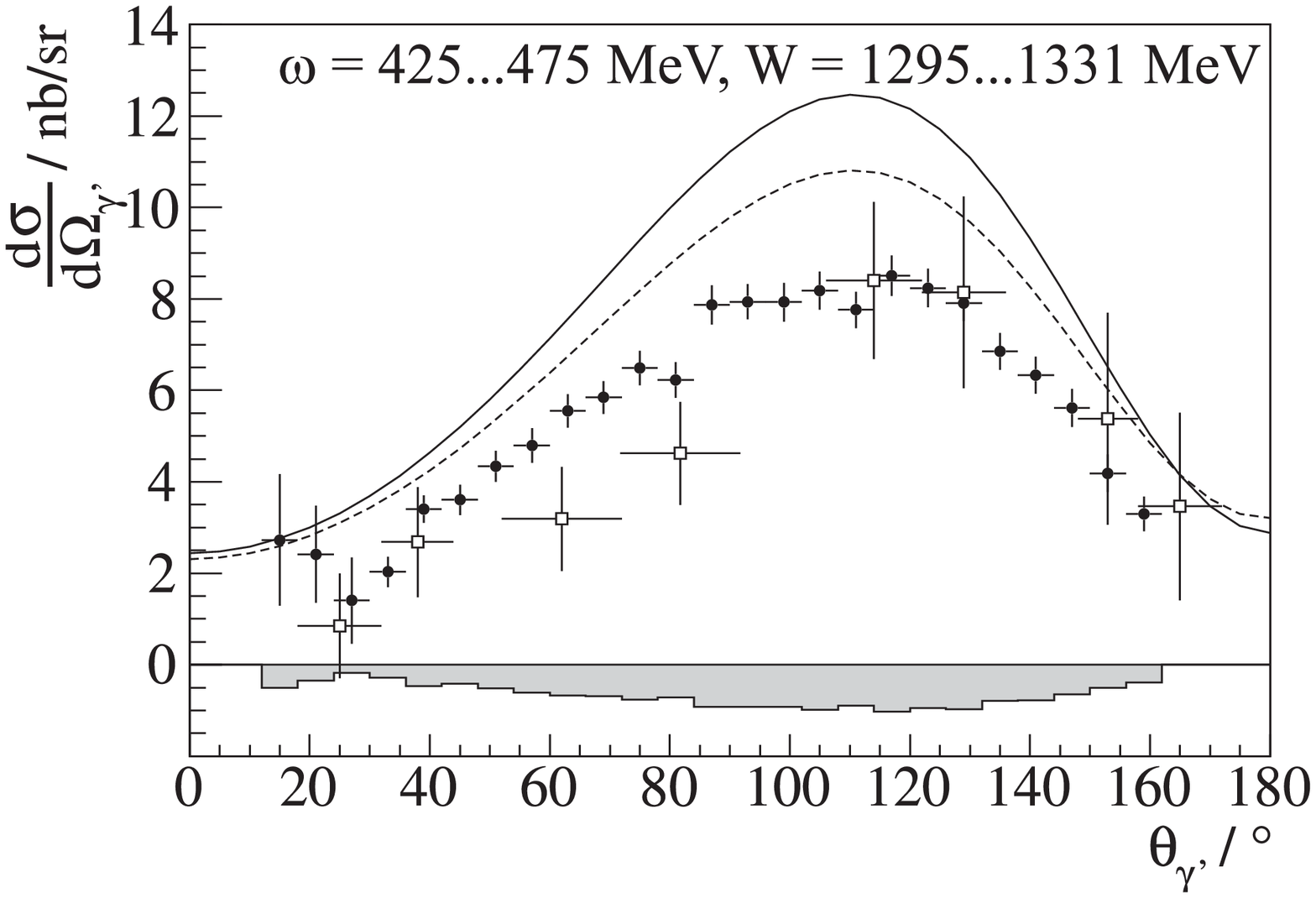}
}

\resizebox{0.99999\textwidth}{!}
{
  \includegraphics{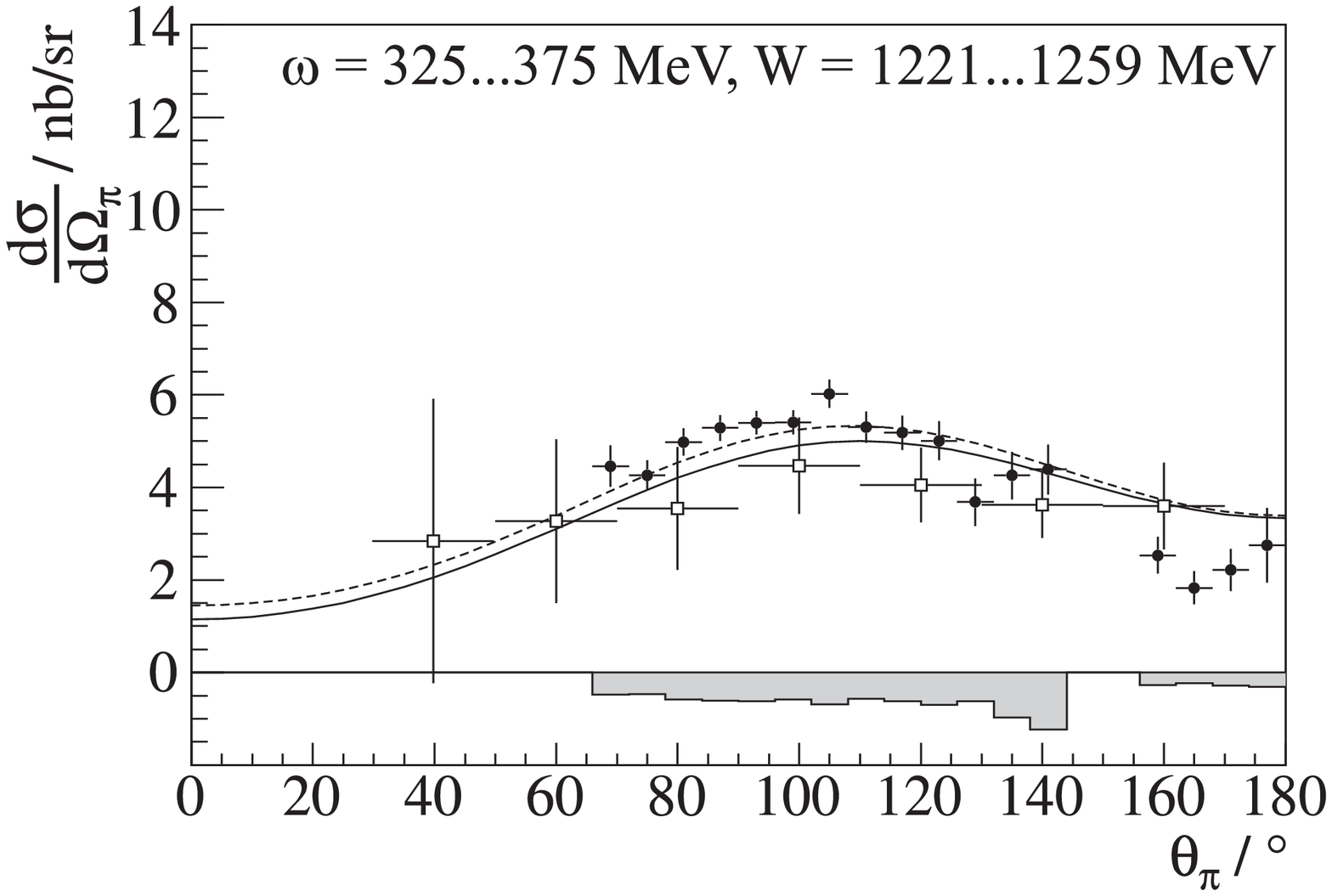}
  \includegraphics{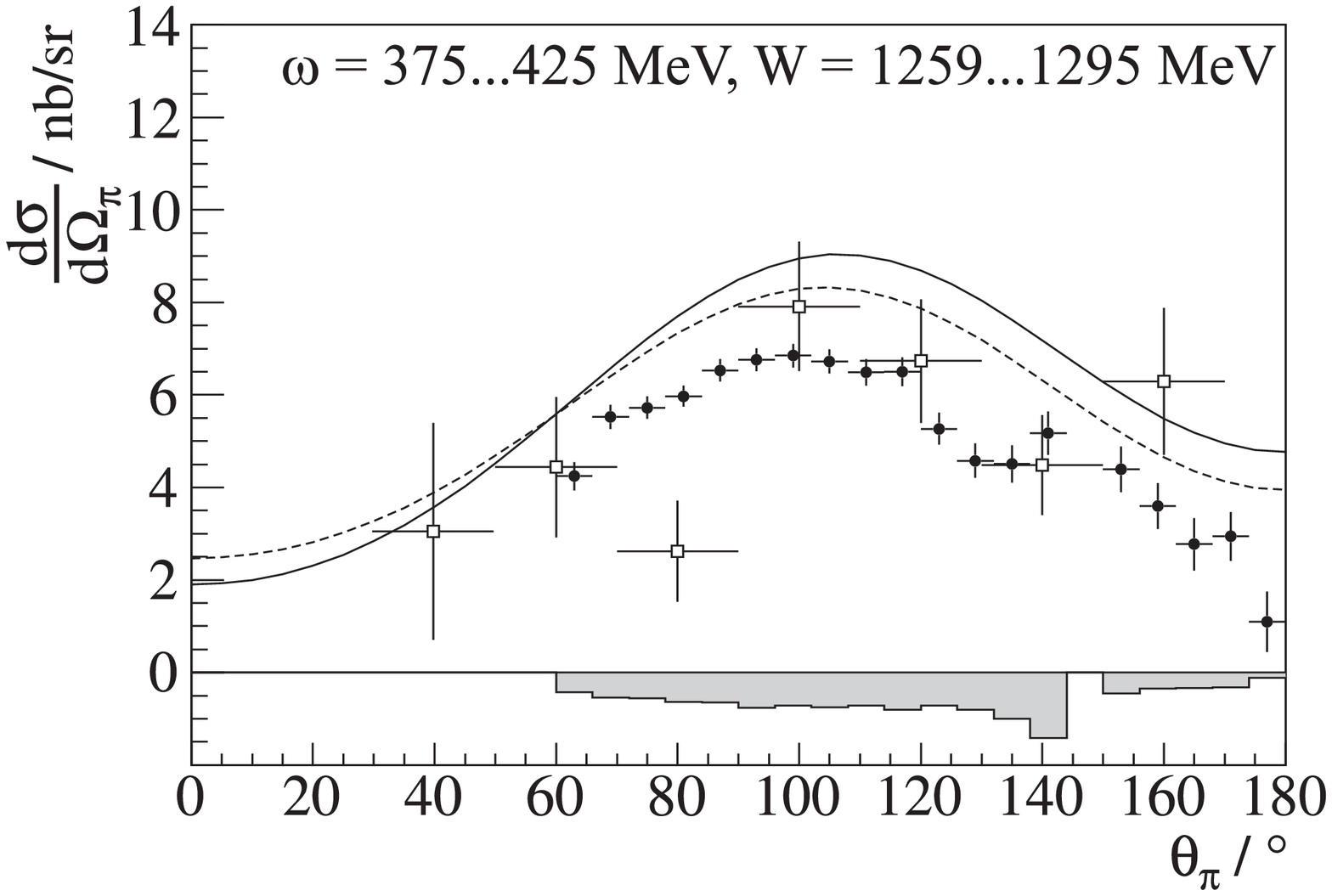}
  \includegraphics{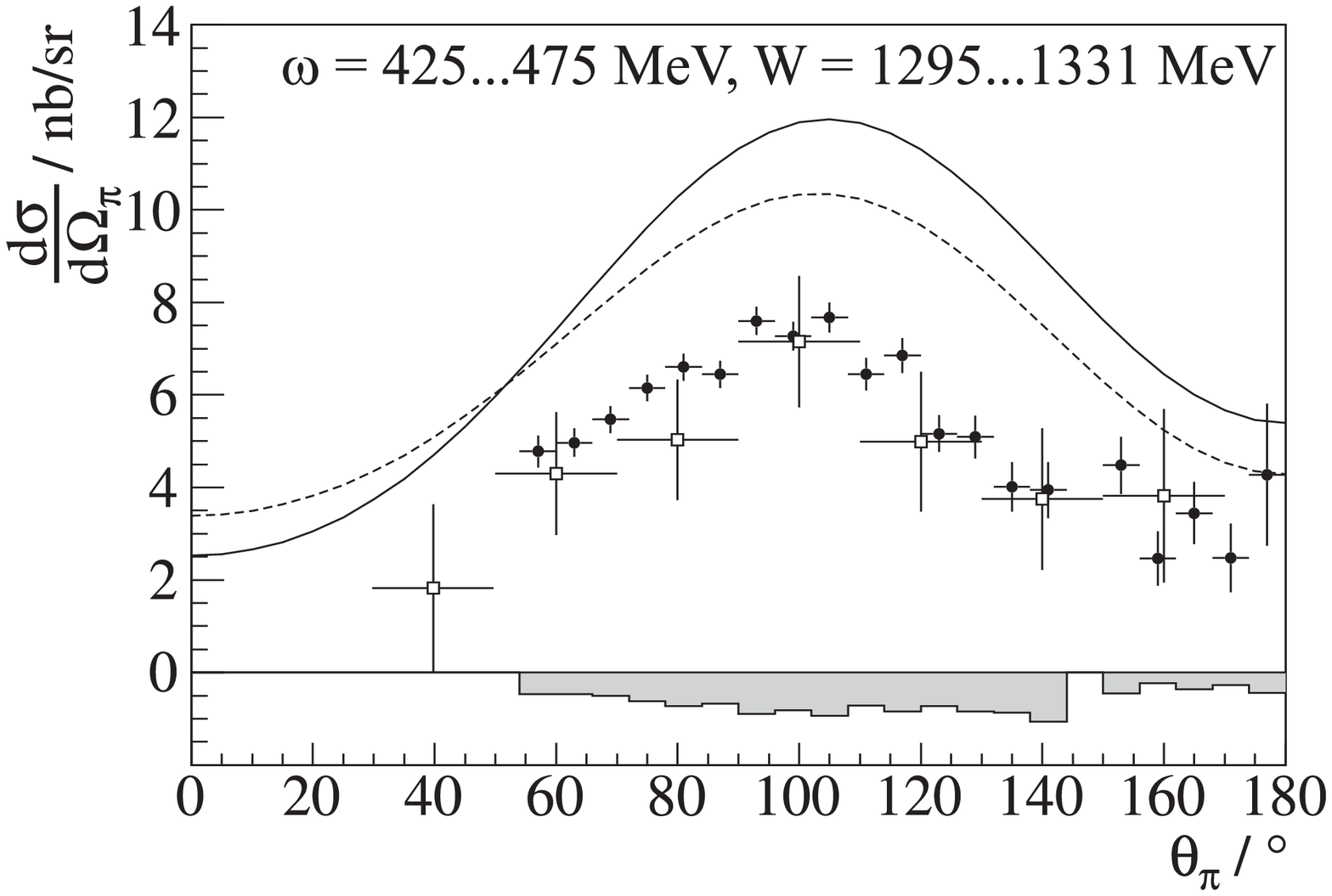}
}
\end{center}
\caption{Differential cross sections at different ranges for beam energy
         $\omega$ and total c.m. energy $W$, respectively. Black points
         represent Crystal Ball / TAPS results, white squares are results from
         ref. \cite{MDMExp}. Error bars denote statistical errors, grey shaded
         bands show absolute systematic uncertainties. Black lines are
         theoretical predictions (using $\kappa_{\Delta^+} = 2.6$) of the
         unitary model from ref. \cite{Unitary} (dashed line) and the
         $\chi$EFT calculation from ref. \cite{ChEFT2} (solid line). Top:
         Energy differential cross section
         $\mathrm{d}\sigma / \mathrm{d}\omega^\prime$ for the emitted photon
         $\gamma^\prime$. Middle: Angular differential cross section 
         $\mathrm{d}\sigma / \mathrm{d}\Omega_{\gamma^\prime}$ for the emitted
         photon $\gamma^\prime$. Bottom: Angular differential cross section 
         $\mathrm{d}\sigma / \mathrm{d}\Omega_\pi$ for the reconstructed
         $\pi^0$ meson.}
\label{fig_dsdX}
\end{figure*}

\section{Results and discussion}
\label{sec_results}

\subsection{\boldmath Single and double $\pi^0$ photoproduction\unboldmath}

Before discussing our new experimental results for radiative $\pi^0$
photoproduction we also present total cross sections for
$\gamma p \rightarrow p \pi^0$ and $\gamma p \rightarrow p \pi^0 \pi^0$
reactions obtained with the data analysis as described in sect.
\ref{sec_analysis}. Figure \ref{fig_totalppi0} shows the total cross sections
for single and double $\pi^0$ production in the energy range from the
particular thresholds up to a beam energy of 800~MeV. For
$\gamma p \rightarrow p \pi^0$ the new Crystal Ball / TAPS results show good
agreement with recent calculations from MAID \cite{MAID1,MAID2} and SAID
\cite{SAID} multipole analyses that reproduce all recent data for differential
and total $\pi^0$ photoproduction cross sections up to 800~MeV. Results for
double $\pi^0$ production are compared to two previous measurements from refs.
\cite{DblPion1a} and \cite{DblPion2} with the TAPS and DAPHNE detectors,
respectively. The overall agreement with both older data sets is rather good.
The small differences are attributed to the phase-space-based efficiency
corrections and covered by the systematic uncertainties. These are estimated
to be 10\% for the previous TAPS data, 4\% for the DAPHNE measurement and 7\%
for our new results.

In addition, fig. \ref{fig_sgldsdO} shows the differential cross section
$\mathrm d\sigma/\mathrm d \Omega_\pi$ for both inclusive and exclusive
measurements of $\gamma p \rightarrow p \pi^0$. In the inclusive analysis, as
described above, only the $\pi^0$ has been reconstructed from its decay into
two photons, while for the exclusive case the recoil proton has also been
detected and identified. This limits the angular acceptance due to detection
limitations for the proton (\textit{e.g.} a minimum kinetic energy of
$T_p \simeq 50$~MeV in the lab frame) to approximately $\theta_\pi = 70^\circ$
to $140^\circ$  and $\theta_\pi > 155^\circ$, where protons are detected in
the Crystal Ball and TAPS, respectively. The acceptance gap between both
detectors is caused mainly by additional material of the inner detector
readout electronics, which blocks the lab frame range from about
$\theta = 10^\circ$ to $20^\circ$ for charged particles. As this gives an
inhomogeneous and complicated acceptance, any remaining charged particles
within this angular range are discarded in the analysis. Both the inclusive
and the exclusive results agree rather well with each other and also with a
previous TAPS / A2 measurement of $\gamma p \rightarrow p \pi^0$ from ref.
\cite{SglPion1} except for $\pi^0$ polar angles around
$\theta_\pi = 130^\circ$. These deviations are attributed to inhomogeneities
in the proton acceptance resulting from the complicated geometry of the
Crystal Ball beam tunnel region. The total discrepancy between inclusive and
exclusive measurements is about 4\%, which is taken into account as an
additional systematic uncertainty for the proton reconstruction efficiency.
In general, the good agreement of our results with model predictions and
previous experiments for single and double $\pi^0$ production indicates that
a reasonable understanding of the main background contributions to
$\gamma p \rightarrow p \pi^0 \gamma^\prime$ reactions has been achieved.

\subsection{\boldmath Radiative $\pi^0$ photoproduction\unboldmath}

For radiative $\pi^0$ production in the 
$\gamma p \rightarrow p \pi^0 \gamma^\prime$ reaction, fig. \ref{fig_dsdX}
shows energy and angular differential cross sections for the emitted photon
$\gamma^\prime$ as well as angular differential cross sections for the $\pi^0$
meson. These cross sections have been determined for three different beam
energy bins, corresponding to c.m. energies $W=\sqrt{s}$ starting close to the
$\Delta^+(1232)$ resonance position and increasing to 100~MeV above. All
cross sections that are not differential in $\gamma^\prime$ energy are
integrated over $\omega^\prime > 30$~MeV. Systematic uncertainties that are
shown as shaded error bands in fig. \ref{fig_dsdX} include uncertainties in
absolute normalisation, namely 5\% for photon flux and 1.8\% for target
density, as well as uncertainties in efficiency correction. This covers
discrepancies between simulated and experimental kinematic distributions,
giving an energy- and angular-dependent contribution to the acceptance
uncertainty which is, on average, around 7\%. Furthermore, a 4\% uncertainty
for proton reconstruction in exclusive measurements is taken into account, as
described above. Finally, for subtraction of single and double $\pi^0$
background a systematic uncertainty of 7.5\% in the number of subtracted
events is assumed. Quadratic addition of all contributions then results in a
mean value of aproximately 10\% for the systematic uncertainties on the
measured cross sections for $\gamma p \rightarrow p \pi^0 \gamma^\prime$.

The energy differential cross sections
$\mathrm{d}\sigma / \mathrm{d}\omega^\prime$ in the upper panels of fig.
\ref{fig_dsdX} are dominated by a $1/\omega^\prime$ behaviour as expected from
bremsstrahlung contributions of the initial and final protons, although,
especially in the highest beam energy interval, this shape is slightly
modified by a broad distribution of emitted photon energies around
$\omega^\prime = 80$~MeV, caused by interference between proton
bremsstrahlung and the $\Delta$-resonant process. Both theoretical models seem
to overestimate this effect and predict a pronounced peak in this energy
range, which is not clearly visible in the experimental results. Compared to
the existing data from ref. \cite{MDMExp} the agreement is good, except for
the low-energy region around emitted photon energies of
$\omega^\prime = 40$~MeV, where the previous measurement resulted in
significantly smaller cross sections.

The angular differential cross sections
$\mathrm{d}\sigma / \mathrm{d}\Omega_{\gamma^\prime}$ show a broad
distribution with a maximum around photon polar angles of about $110^\circ$
(see middle panels of fig. \ref{fig_dsdX}). This is caused by interference
bet\-ween bremsstrahlung from the initial and final protons and the
$\Delta$-resonant mechanism, while a pure $\Delta$-resonant process would
produce an angular distribution peaking at $90^\circ$. Both model calculations
in general reproduce the shape of the angular distributions but overestimate
the absolute cross sections at higher beam energies, while there is reasonable
agreement between the old TAPS / A2 measurement and our new data.

Angular distributions $\mathrm{d}\sigma / \mathrm{d}\Omega_{\pi}$ for the
$\pi^0$ meson have been determined for polar angles in the range from
$\theta_\pi = 60^\circ$ to $140^\circ$ and $\theta_\pi > 150^\circ$ (see lower
panels of fig. \ref{fig_dsdX}). As in the analysis of
$\gamma p \rightarrow p \pi^0 \gamma^\prime$ reactions any charged particles
within the laboratory polar angle range between $10^\circ$ and $20^\circ$ are
discarded due to the complicated acceptance in this region. This is comparable
to the single $\pi^0$ case (see fig. \ref{fig_sgldsdO}), as the kinematics for
$\gamma p \rightarrow p \pi^0 \gamma^\prime$ are very similar to
$\gamma p \rightarrow p \pi^0$ especially for soft photons $\gamma^\prime$
resulting from the dominating proton bremsstrahlung. Also for this observable,
the cross section shape is reproduced quite well by theoretical calculations,
but the the discrepancies in the absolute values increase more and more with
rising beam energy.

\begin{figure}
\begin{center}
\resizebox{0.49\textwidth}{!}
{
  \includegraphics{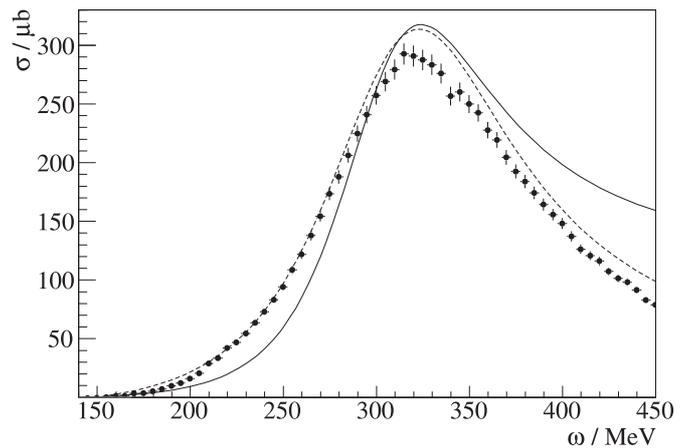}
}
\end{center}
\caption{Total cross section for $\gamma p \rightarrow p \pi^0$ in the energy
         region of the $\Delta^+(1232)$. Data points represent
         Crystal Ball / TAPS results, error bars include both statistical and
         systematic uncertainties. Dashed and solid lines show predictions
         from the unitarised effective Lagrangian model \cite{Unitary} and the
         $\chi$EFT calculation \cite{ChEFT2}, respectively.}
\label{fig_sgltotal}
\end{figure}

These discrepancies in the cross section predictions are mainly attributed to
inaccuracies in the description of the non-radiative
$\gamma p \rightarrow p \pi^0$ reaction (see fig. \ref{fig_sgltotal}), which
is connected to the $\gamma p \rightarrow p \pi^0 \gamma^\prime$ process by
the low-energy theorem for $\omega^\prime \rightarrow 0$. This is particularly
important for the $\chi$EFT calculation from ref. \cite{ChEFT2}, that does not
include any vector meson ($\rho$, $\omega$) exchange mechanisms. These
processes, however, dominate the high-energy behaviour of
$\gamma p \rightarrow p \pi^0$, where the $\Delta$ excitation alone yields
cross sections that are too large and are subsequently reduced by the addition
of $\rho, \omega$ exchange mechanisms. Some of these model dependencies can be
avoided by construction of a suitable observable for
$\gamma p \rightarrow p \pi^0 \gamma^\prime$ that does not depend on details
of the $\gamma p \rightarrow p \pi^0$ reaction description.

\begin{figure*}
\begin{center}
\resizebox{0.99999\textwidth}{!}
{
  \includegraphics{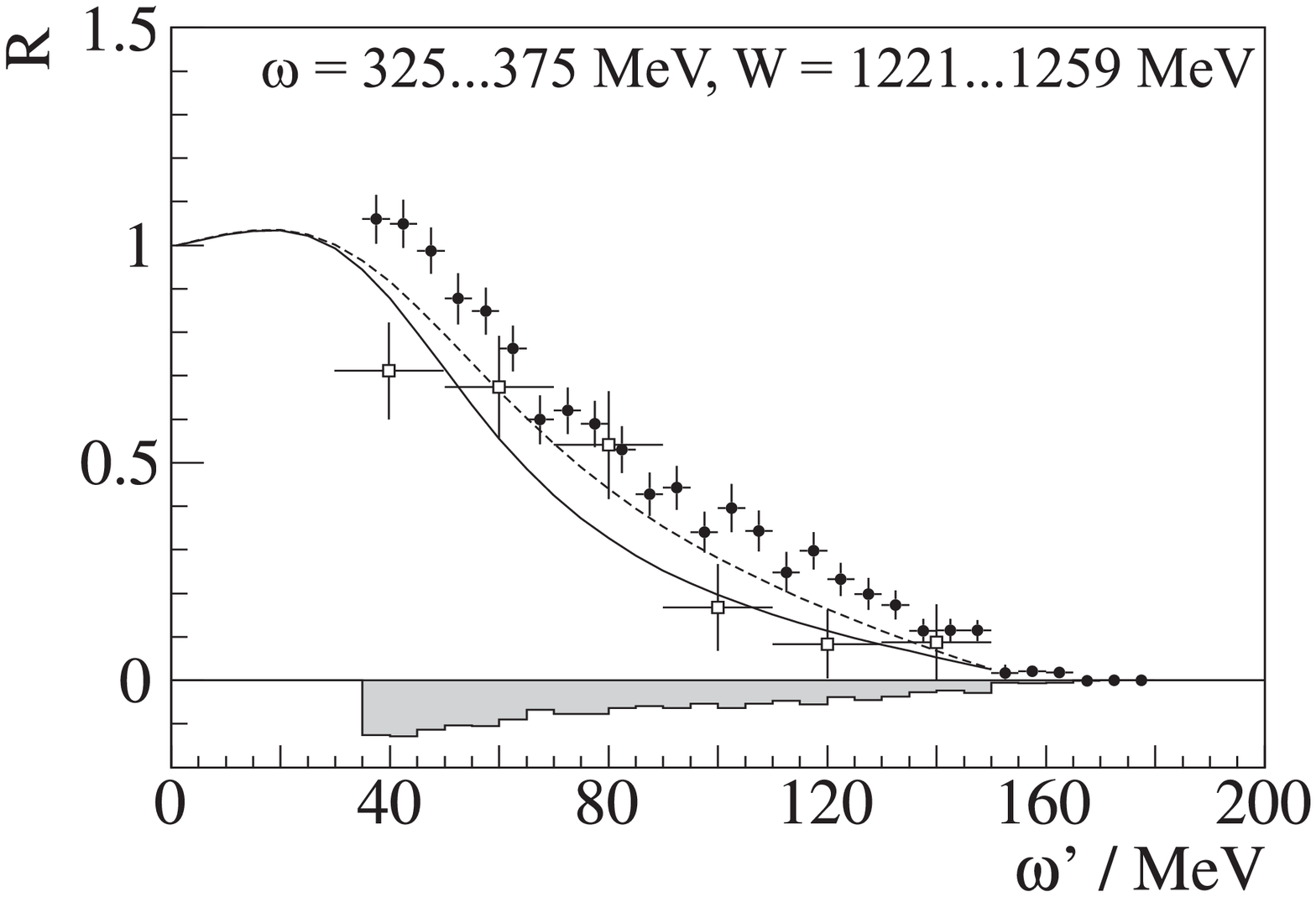}
  \includegraphics{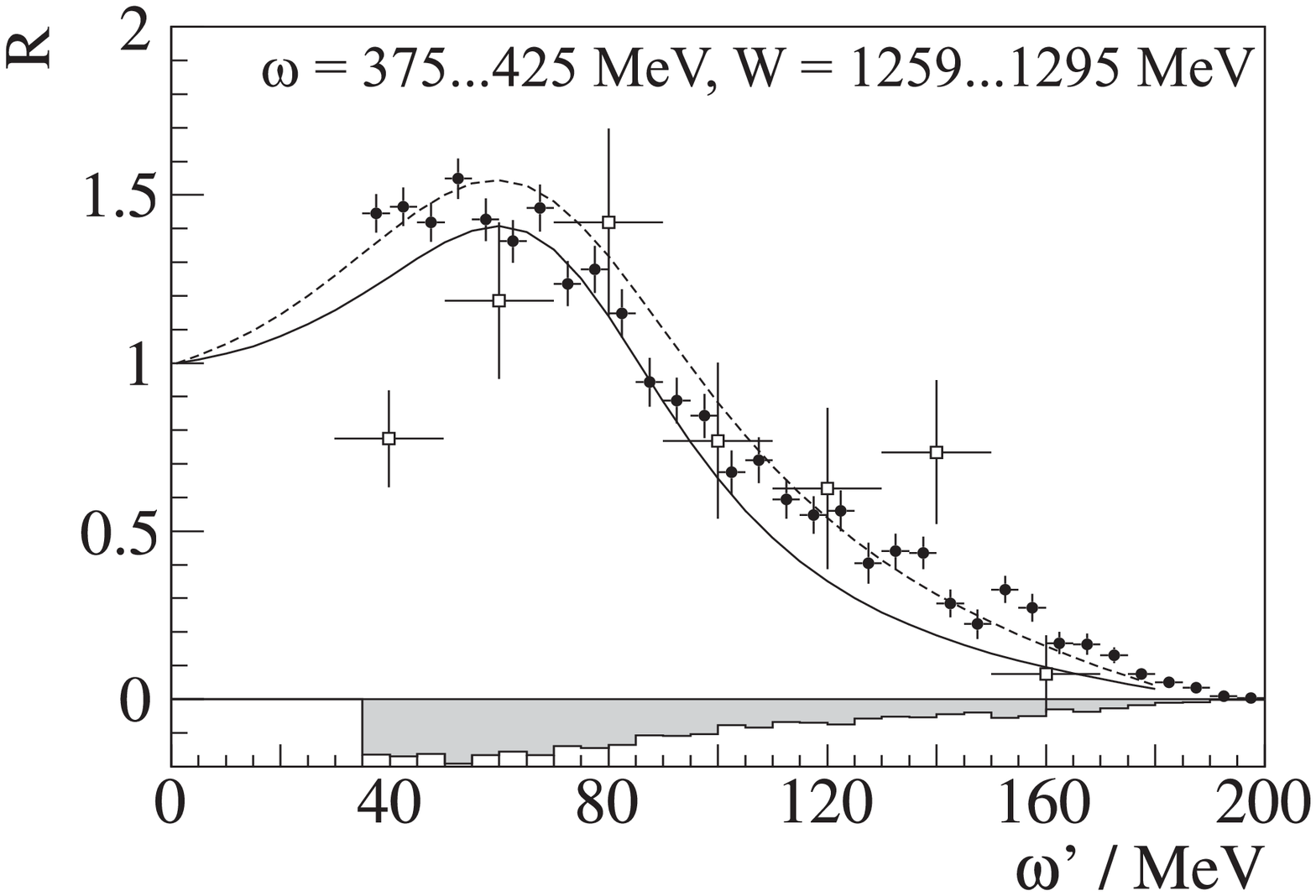}
  \includegraphics{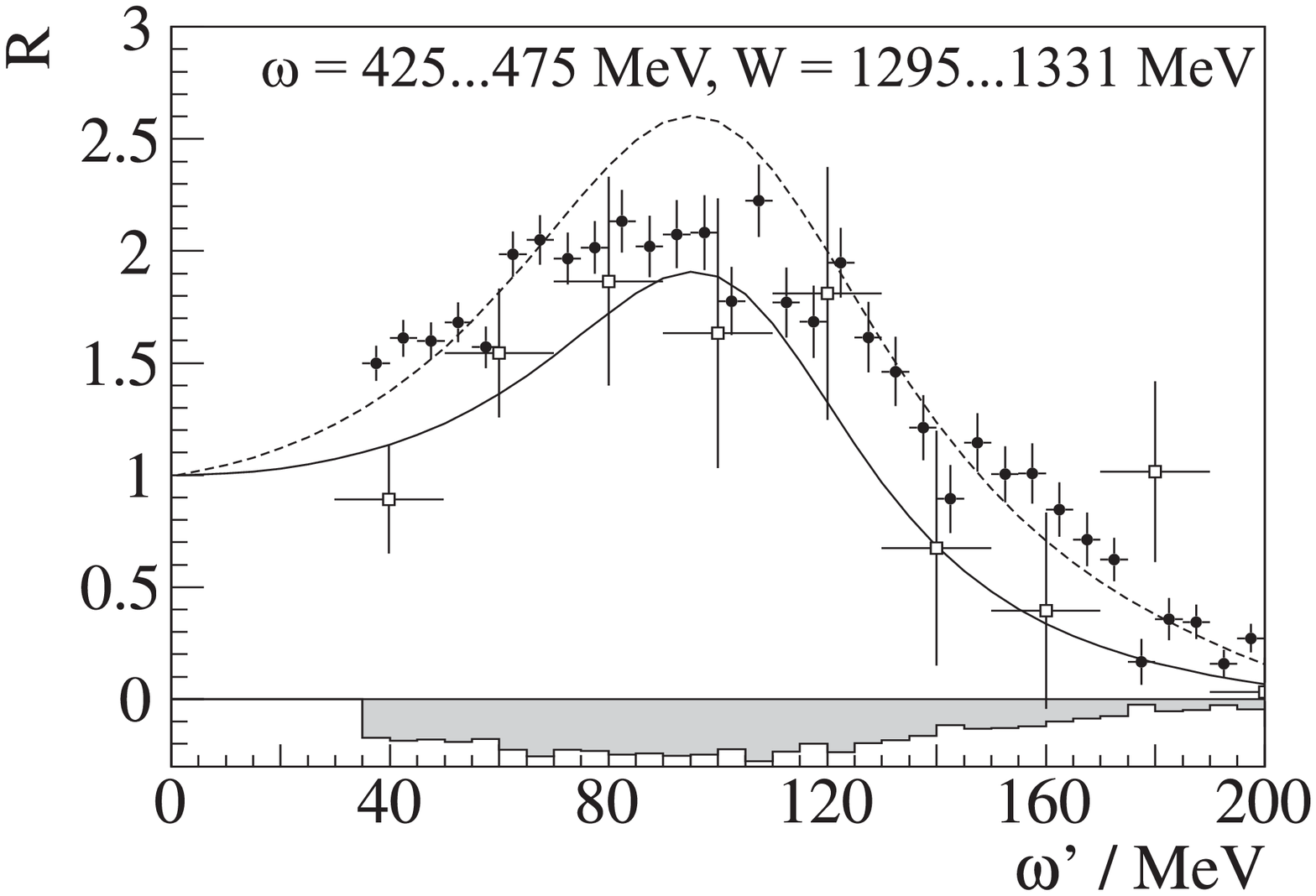}
}
\end{center}
\caption{Cross section ratio $R$ at different ranges for beam energy $\omega$
         and total c.m. energy $W$, respectively. Black points represent
         Crystal Ball / TAPS results, white squares are results from ref.
         \cite{MDMExp}. Error bars denote statistical errors, grey shaded
         bands show absolute systematic uncertainties. Black lines are
         theoretical predictions (using $\kappa_{\Delta^+} = 2.6$) of the
         unitary model from ref. \cite{Unitary} (dashed line) and the
         $\chi$EFT calculation from ref. \cite{ChEFT2} (solid line).}
\label{fig_ratio}
\end{figure*}

In the soft-photon limit ($\omega^\prime \rightarrow 0$) radiative $\pi^0$
production is completely determined by bremsstrahlung processes from the
initial and final protons and gauge invariance provides a model-independent
relation between $\gamma p \rightarrow p \pi^0 \gamma^\prime$ and
non-radiative $\gamma p \rightarrow p \pi^0$ reactions. As shown in ref.
\cite{SoftPhoton}, in this soft-photon limit the three-fold differential cross
section, after integration over the outgoing photon angles, is given by
\begin{equation}
\frac{\mathrm{d}^3\sigma}{\mathrm{d}\omega^\prime \mathrm{d}\Omega_\pi}
\stackrel{\omega^\prime \rightarrow 0}{\longrightarrow}
\frac{1}{\omega^\prime} \cdot \frac{e^2}{2\pi^2}\cdot W(v)\cdot 
\frac{\mathrm{d}\sigma}{\mathrm{d}\Omega_\pi}
\end{equation}
with $\mathrm{d}\sigma/\mathrm{d}\Omega_\pi$ being the differential cross
section for the $\gamma p \rightarrow p \pi^0$ process, and an angular weight
function
\begin{equation}
W(v) = \left(\frac{v^2+1}{2v}\right)\cdot\ln\left(\frac{v+1}{v-1}\right) - 1
\end{equation}
with
\begin{equation}
v = \sqrt{1 - \frac{4m_p}{(k-q)^2}}
\end{equation}
where $k$, $q$ denote the 4-momenta of the beam photon and the $\pi^0$,
respectively. Integration over the meson angles yields an energy distribution
\begin{equation}
\label{eqn_lowenergy}
\frac{\mathrm{d}\sigma}{\mathrm{d}\omega^\prime}
\stackrel{\omega^\prime \rightarrow 0}{\longrightarrow}
\frac{1}{\omega^\prime} \cdot \sigma_\pi
\end{equation}
with a weight-averaged total cross section
\begin{equation}\label{frm_sigmapi}
\sigma_\pi = \frac{e^2}{2\pi^2}\ \int \mathrm{d}\Omega_\pi \; W(v)\cdot 
\frac{\mathrm{d}\sigma}{\mathrm{d}\Omega_\pi}
\end{equation}
for the $\gamma p \rightarrow p \pi^0$ reaction.
From the low-energy theorem of eq. (\ref{eqn_lowenergy}), as derived in
appendix B of ref. \cite{SoftPhoton}, the cross section ratio
\begin{equation}\label{frm_ratio}
R = \frac{1}{\sigma_\pi}\cdot \omega^\prime \cdot 
\frac{\mathrm{d}\sigma}{\mathrm{d}\omega^\prime}
\end{equation}
is defined, with the soft-photon limit value $R \rightarrow 1$ for vanishing
photon energies $\omega^\prime \rightarrow 0$.

\begin{table}
\caption{Weight-averaged total cross sections $\sigma_\pi$. Theoretical
         predictions are from the unitary model \cite{Unitary} and the
         $\chi$EFT calculation \cite{ChEFT2}. Experimental values refer to
         Crystal Ball / TAPS results with statistical and systematic errors.}
\label{tab_sigmapi}
\begin{center}
\begin{tabular}{lccc}
\hline\noalign{\smallskip}
Beam            & \multicolumn{3}{c}{Weight-averaged cross
                                     section $\sigma_\pi$}   \\
energy $\omega$ & unitary & $\chi$EFT & experimental         \\
\noalign{\smallskip}\hline\noalign{\smallskip}
350~MeV & 56.97~nb & 60.51~nb & $56.18 \pm 0.06 \pm 2.26$~nb \\
400~MeV & 42.17~nb & 51.82~nb & $40.59 \pm 0.06 \pm 1.64$~nb \\
450~MeV & 32.31~nb & 49.47~nb & $25.91 \pm 0.07 \pm 1.04$~nb \\
\noalign{\smallskip}\hline
\end{tabular}
\end{center}
\end{table}

This ratio is less sensitive not only to experimental uncertainties
(\textit{e.g.} photon flux, target density) which contribute to both
$\mathrm{d}\sigma/\mathrm{d}\omega^\prime$ and $\sigma_\pi$ and hence cancel
out in $R$, but also to inaccuracies in theoretical model calculations of
$\gamma p \rightarrow p \pi^0$, as this observable depends only on the
deviation from the soft-photon behaviour. Quantitative effects of these model
dependencies can be estimated through the weight-averaged total cross section
$\sigma_\pi$ from eq. (\ref{frm_sigmapi}) for the 
$\gamma p \rightarrow p \pi^0$ reaction. Table \ref{tab_sigmapi} compares
predictions for $\sigma_\pi$ from both the unitary model \cite{Unitary} and
the $\chi$EFT calculation \cite{ChEFT2} with experimental Crystal Ball / TAPS
results obtained from an analysis of $\gamma p \rightarrow p \pi^0$ at three
different beam energies $\omega$ in the considered energy range. While the
unitarised effective Lagrangian framework \cite{Unitary} gives good agreement
with experimental results at least in the low and medium photon energy range,
the $\chi$EFT calculation \cite{ChEFT2} shows discrepancies of about 10\%
already at low photon energies, increasing to nearly a factor of two at the
highest beam energy. As mentioned above, this is due to reaction mechanisms
not included in the $\chi$EFT expansion and, therefore, a well-understood
limitation, which does not affect the description of $\Delta$-resonant
processes.

With these results for the weight-averaged cross section $\sigma_\pi$ for the
non-radiative $\gamma p \rightarrow p \pi^0$ reaction and the energy
differential cross sections $\mathrm{d}\sigma/\mathrm{d}\omega^\prime$ from
fig. \ref{fig_dsdX}, cross section ratios $R$ have been derived according to
eq. (\ref{frm_ratio}) for the three different beam energy ranges (see fig.
\ref{fig_ratio}). Also for this observable, the same value for the anomalous
magnetic moment $\kappa_{\Delta^+}$ results in significant discrepancies
between the unitary and $\chi$EFT predictions. This shows that there is still
a remaining model dependence in the description of
$\gamma p \rightarrow p \pi^0 \gamma^\prime$, which does not cancel in $R$
and, therefore, is not related to details of the 
$\gamma p \rightarrow p \pi^0$ reaction model. Such model dependencies,
however, limit the accuracy of any values for $\kappa_{\Delta^+}$ extracted
from experi\-mental results.

\begin{figure*}
\begin{center}
\resizebox{0.99999\textwidth}{!}
{
  \includegraphics{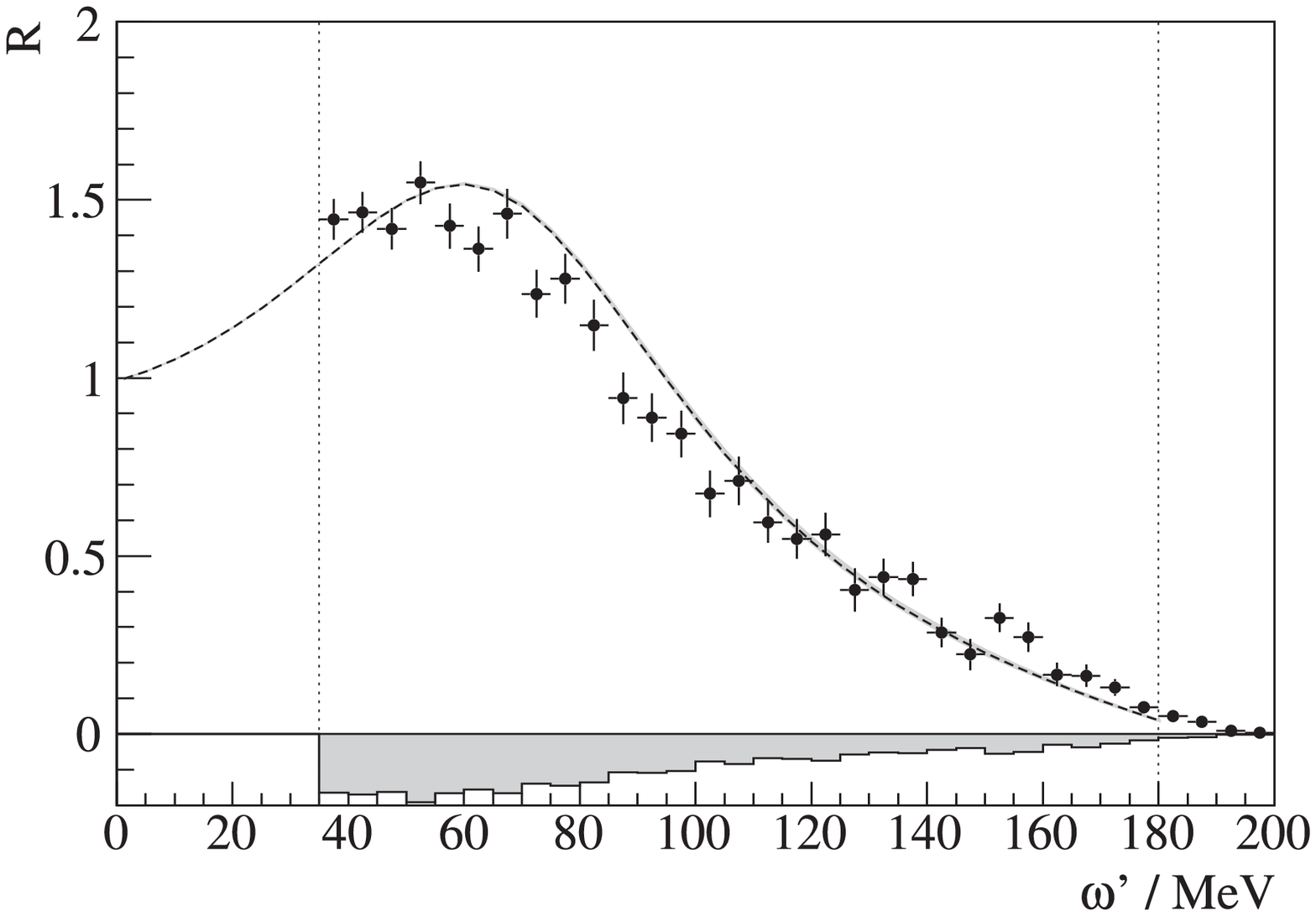}
  \includegraphics{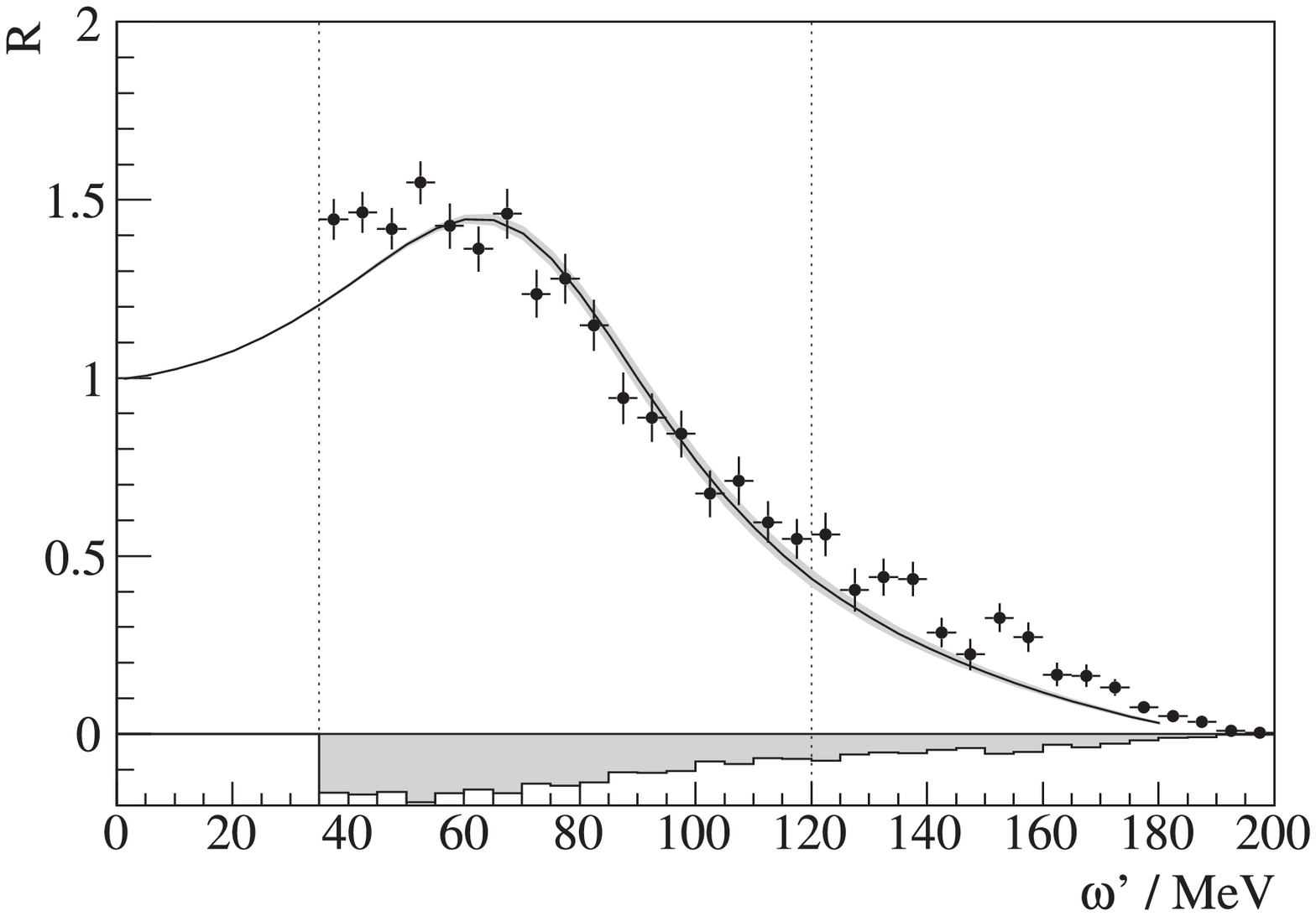}
}
\end{center}
\caption{Cross section ratio $R$ for beam energies from
         $\omega=375$ to $425$~MeV. The dashed and solid lines represent the
         best $\chi^2$ fit for the unitary model (left) and the $\chi$EFT
         calculation (right) with $\kappa_{\Delta^+}$ as free parameter. Grey
         shaded bands show the variations of the model predictions within the
         statistical fit errors for $\kappa_{\Delta^+}$, vertical lines
         indicate the range of emitted photon energies used for fitting.}
\label{fig_fits}
\end{figure*}

The following considerations and fits to experimental results will concentrate
on the medium beam energy range around $\omega = 400$~MeV, as this is a
compromise between sensitivity to $\kappa_{\Delta^+}$, which is fairly low at
lower beam energies, and the agreement between experimental data and
theoretical calculations, which gets worse with rising beam energy. Such a
procedure that uses only a limi\-ted part of our experimental results is not
suitable for an unambiguous determination of the $\Delta^+(1232)$ magnetic
dipole moment, but will still give an impression of the statistical and
systematic precision for $\mu_{\Delta^+}$ that might be achieved if improved
theoretical models become available. The left panel of fig. \ref{fig_fits}
shows a $\chi^2$ fit of unitary model predictions to our new results for $R$
for emitted photon energies up to the kinematic limit at
$\omega^\prime \simeq 180$~MeV with the anomalous magnetic moment
$\kappa_{\Delta^+}$ as free parameter. This fit gives a result of
$\kappa_{\Delta^+} = 2.79^{+0.39}_{-0.44}(\mbox{stat})\pm0.39(\mbox{syst})$,
but the obtained fit quality of $\chi^2/\mbox{ndf} = 2.94$ ($\mbox{ndf} = 28$)
indicates that, even with the restriction to beam energies around
$\omega = 400$~MeV, the agreement between experimental data points and
theo\-retical model calculations in the unitarised effective Lagrangian
framework \cite{Unitary} is not sufficient for a reliable extraction of the
$\Delta^+(1232)$ magnetic dipole moment. As already mentioned in the
discussion of the energy differen\-ti\-al cross sections
$\mathrm d\sigma/\mathrm d \omega^\prime$ the model overestimates a peak
structure for energies around $\omega^\prime = 60$ to $100$~MeV and does not
fully reproduce the shape of the experimental distribution.

In the case of the $\chi$EFT calculation (see right panel of fig.
\ref{fig_fits}) it is reasonable to limit the fit to emitted photon energies
up to $\omega^\prime = 120$~MeV \cite{PrivComm1}, which yields a value of
$\kappa_{\Delta^+} = 3.95^{+0.18}_{-0.20}(\mbox{stat})\pm0.85(\mbox{syst})$
at $\chi^2/\mbox{ndf} = 2.72$ ($\mbox{ndf} = 16$). This restriction to
low-energy photons $\gamma^\prime$ is motivated by the expansion schemes used
in the $\chi$EFT framework, which require the initial beam photon to be in the
order of the $\Delta^+(1232)$ excitation energy, while the emitted photon
$\gamma^\prime$ has to be soft. Applicability of this model is, therefore,
limited to beam energies not too far away from the resonance point (which
gives another reason for ruling out the higher beam energy bin at
$\omega = 450$~MeV), while the low-energy expansion in $\omega^\prime$ becomes
problematic for $\gamma^\prime$ energies near the kinematic limit. However,
also at low photon energies around $\omega^\prime = 40$~MeV there still
remains a discrepancy bet\-ween the experimental results and the $\chi$EFT
calculation. Thus, at the moment none of these theoretical descriptions seems
to give a precise description of the experimental results that would be needed
for a reliable extraction of the anomalous magnetic moment
$\kappa_{\Delta^+}$. Furthermore, the fit results for $\kappa_{\Delta^+}$
obtained from both models are not compatible with each other.

\section{Conclusion and outlook}
\label{sec_outlook}

A new measurement of radiative $\pi^0$ photoproduction
$\gamma p \rightarrow p \pi^0 \gamma^\prime$ in the $\Delta^+(1232)$ energy
region has been performed. The data were obtained with the Crystal Ball / TAPS
detectors using an energy-tagged photon beam produced at the electron
accelerator facility MAMI-B. Compared to the pioneering TAPS / A2 experiment
in ref. \cite{MDMExp}, from which the $\Delta^+(1232)$ magnetic dipole moment
$\mu_{\Delta^+}$ was extracted for the first time, a considerable improvement
on statistics by nearly a factor of 60 has been achieved. This new experiment
yielded differential cross sections for both the emitted photon
$\gamma^\prime$ and the $\pi^0$ with high resolutions in polar angles
$\theta_{\gamma^\prime}$ and $\theta_\pi$ and emitted photon energy
$\omega^\prime$. These cross sections show a reasonable agreement with the
previous experimental result, but cannot be fully reproduced by different
theoretical models \cite{Unitary,ChEFT2} for
$\gamma p \rightarrow N \pi \gamma^\prime$ reactions.

Part of this discrepancy vanishes if the cross section ratio $R$ from eq. 
(\ref{frm_ratio}) is evaluated, as this eliminates some of the model
dependencies, but also with this observable a reliable and precise extraction
of the magnetic dipole moment $\mu_{\Delta^+}$ seems to be prevented by
limitations in the applicability of the models and their inability to describe
our experimental data accurately. Thus, substantial progress on the model
descriptions is required in order to fully exploit the present experimental
results, as the large improvement in the quality of this data compared to the
older measurement of radiative $\pi^0$ photoproduction reveals shortcomings in
the models that previously were not apparent due to the limited statistics in
ref. \cite{MDMExp}. In this context already different approaches for
the on-shell propagator in the $\Delta\rightarrow\Delta\gamma^\prime$ process,
as they have been recently studied for three models in ref. \cite{DeltaProp},
turn out to have significant influence on both shape and magnitude of the
theoretical predictions for $\gamma p \rightarrow p\pi^0 \gamma^\prime$ cross
sections.

In the meantime also several other observables have been suggested that show
a larger and more direct dependence on $\mu_{\Delta^+}$, \textit{i.e.} not
relying on interference effects between the $\Delta$-resonant process and
bremsstrahlung contributions. For example, ref. \cite{ChEFT2} proposes to
measure helicity asymmetries $\Sigma_\mathrm{circ}$ with a circularly
polarised photon beam. These helicity asymmetries are nonzero only for
three-body final states and vanish exactly in the the soft-photon limit
$\omega^\prime \rightarrow 0$, as $\gamma p \rightarrow N \pi \gamma^\prime$
processes reduce to $\gamma p \rightarrow N \pi$ two-body reactions.
Furthermore, $\Sigma_\mathrm{circ}$ shows a linear dependence on
$\mu_{\Delta^+}$, where it should be noted that this is a model-independent
feature because of the low-energy theorem for
$\gamma p \rightarrow N \pi \gamma^\prime$. This model-independent
determination of $\mu_{\Delta^+}$ would require a measurement with a high
degree of circular photon beam polarisation in the energy region around the
$\Delta^+(1232)$ resonance point. However, as the predicted magnitude of these
helicity asymmetries is in the order of only about 1\%, such a measurement
would require a further major improvement in statistics which seems currently
out of reach.

\section{Acknowledgments}
\label{sec_acknow}
It is a pleasure to acknowledge useful discussions with V. Pascalutsa and
M. Vanderhaeghen. The authors thank the MAMI accelerator group and operators
for their excellent support. This work was supported by Deutsche 
Forschungsgemeinschaft (SFB 443, SFB/TR 16), DFG-RFBR (Grant No. 05-02-04014),
European Com\-munity-Research Infrastructure Activity under FP6 ``Structuring
the European Research Area'' programme (HadronPhysics, Contract No.
RII3-CT-2004-506078), Schweizeri\-scher Nationalfonds, NSERC (Canada), UK
EPSRC and STFC, US DOE and US NSF. We thank the undergraduate students of
Mount Allison University and The George Washington University for their
assistance.

\end{document}